\documentclass{aa}  
\usepackage{graphicx}
\usepackage{subcaption}
\usepackage{caption}
\usepackage{changepage}
\usepackage{xcolor}
\usepackage{booktabs}
\usepackage{txfonts}
\usepackage[bookmarks, colorlinks, breaklinks, ]{hyperref} 
\usepackage{natbib}
\hypersetup{linkcolor=blue,citecolor=blue,filecolor=blue,urlcolor=blue} 
\usepackage{hyperref}
\usepackage{textcomp}
\def\sex{{\tt SExtractor}}
\def\psfex{{\tt PSFEx}}
\def\scamp{{\tt scamp}}
\def\swarp{{\tt SWarp}}

\def\astromatic{Astr\textit{O}matic}

\begin{document}

   \title{CHEX-MATE: Multi-probe analysis of Abell 1689}

   \author{L. Chappuis
          \inst{\ref{cea}, \ref{gva}} \and
          D. Eckert
          \inst{\ref{gva}} \and
          M. Sereno
          \inst{\ref{oabo}} \and
          A. Gavidia
          \inst{\ref{caltech}} \and
          J. Sayers
          \inst{\ref{caltech}} \and
          J. Kim
          \inst{\ref{kaist}} \and
          M. Rossetti
          \inst{\ref{iasfmi}} \and
          K. Umetsu
          \inst{\ref{asiaa}} \and
          H. Saxena
          \inst{\ref{caltech}} \and
          I. Bartalucci
          \inst{\ref{iasfmi}} \and
          R. Gavazzi
          \inst{\ref{inst1},\ref{inst2}} \and
          A. Rowlands Doblas
          \inst{\ref{gva}} \and
          E. Pointecouteau
          \inst{\ref{irap}} \and
          S. Ettori 
          \inst{\ref{oabo},\ref{infn}} \and
          G. W. Pratt
          \inst{\ref{cea}} \and
          H. Bourdin
          \inst{\ref{tv},\ref{infnro}} \and
          R. Cassano
          \inst{\ref{ira}} \and
          F. De Luca
          \inst{\ref{tv},\ref{infnro}} \and
          M. Donahue
          \inst{\ref{umich}} \and
          M. Gaspari
          \inst{\ref{umod}} \and
          F. Gastaldello
          \inst{\ref{iasfmi}} \and
          V. Ghirardini
          \inst{\ref{oabo}} \and
          M. Gitti
          \inst{\ref{ubo},\ref{ira}} \and
          B. Maughan
          \inst{\ref{bristol}} \and
          P. Mazzotta
          \inst{\ref{tv},\ref{infnro}}\and
          F. Oppizzi
          \inst{\ref{infnge}}\and
          E. Rasia
          \inst{\ref{oatri},\ref{ifpu},\ref{mich}}\and
          M. Radovich
          \inst{\ref{oapad}}
          }

   \institute{Université Paris Saclay, IRFU/CEA, 91191, Gif-sur-Yvette, France\label{cea} \\
               \email{loris.chappuis@cea.fr}
               \and
               Department of Astronomy, University of Geneva, Ch. d’Ecogia 16, CH-1290 Versoix, Switzerland\label{gva}
               \and
               INAF, Osservatorio di Astrofisica e Scienza dello Spazio, Via Piero Gobetti 93/3, 40129, Bologna, Italy\label{oabo}
               \and
               California Institute of Technology, 1200 E. California Blvd., MC 367-17, Pasadena, CA, 91125, USA\label{caltech}
               \and
               Korea Advanced Institute of Science and Technology (KAIST), 291 Daehak-ro, Yuseong-gu, Daejeon, 34141, Republic of Korea\label{kaist}
               \and
               INAF, IASF-Milano, via A. Corti 12, 20133, Milano, Italy\label{iasfmi}
               \and
               Academia Sinica Institute of Astronomy and Astrophysics (ASIAA), No. 1, Section 4, Roosevelt Road, Taipei 10617, Taiwan\label{asiaa}
               \and
               Laboratoire d'Astrophysique de Marseille, Aix-Marseille Univ., CNRS, CNES, Marseille, France\label{inst1}
               \and
              Institut d'Astrophysique de Paris, UMR7095 CNRS \& Sorbonne Universit\'e, 98bis Bd Arago, F-75014, Paris\label{inst2}
              \and
              IRAP, CNRS, Université de Toulouse, CNES, UT3-UPS, (Toulouse), France\label{irap}
              \and
              INFN, Sezione di Bologna, viale Berti Pichat 6/2, 40127 Bologna, Italy\label{infn}
              \and
              Dipartimento di Fisica, Universit\`a degli studi di Roma Tor Vergata, Via della Ricerca Scientifica 1, I-00133 Roma, Italy\label{tv}
              \and
              INFN, Sezione di Roma ‘Tor Vergata’, Via della Ricerca Scientifica, 1, 00133, Roma, Italy\label{infnro}
              \and
              INAF - Istituto di Radioastronomia, via P. Gobetti, n.101, I-40129, Bologna, Italy\label{ira}
              \and
              Michigan State University, Physics \& Astronomy Department, 567 Wilson Rd, East Lansing, MI, USA, 48824\label{umich}
              \and
              Department of Physics, Informatics and Mathematics, University of Modena and Reggio Emilia, 41125 Modena, Italy\label{umod}
              \and
              Dipartimento di Fisica e Astronomia (DIFA), Alma Mater Studiorum - Università di Bologna, via Gobetti 93/2, 40129 Bologna, Italy\label{ubo}
              \and
              HH Wills Physics Laboratory, University of Bristol, Bristol, UK\label{bristol}
              \and
              INFN - Sezione di Genova, Via Dodecaneso 33, 16146, Genova, Italy\label{infnge}
              \and
              INAF -- Osservatorio Astronomico di Trieste, via Tiepolo 11, I-34131, Trieste, Italy\label{oatri}
              \and
              IFPU, Institute for Fundamental Physics of the Universe, Via Beirut 2, 34014 Trieste, Italy\label{ifpu}
              \and
              Department of Physics; University of Michigan, Ann Arbor, MI 48109, USA\label{mich}
              \and
              INAF -- Osservatorio astronomico di Padova, vicolo dell’Osservatorio 5, I-35122, Padova, Italy\label{oapad}
            }


 
  \abstract
    {The nature of the elusive dark matter can be probed by comparing the predictions of the cold dark matter framework with the gravitational field of massive galaxy clusters. However, a robust test of dark matter can only be achieved if the systematic uncertainties in the reconstruction of the gravitational potential are minimized. Techniques based on the properties of intracluster gas rely on the assumption that the gas is in hydrostatic equilibrium within the potential well, whereas gravitational lensing is sensitive to projection effects. Here we attempt to minimize systematics in galaxy cluster mass reconstructions by jointly exploiting the weak gravitational lensing signal and the properties of the hot intracluster gas determined from X-ray and millimeter (Sunyaev-Zel'dovich) observations. We construct a model to fit the multi-probe information within a common framework, accounting for non-thermal pressure support and elongation of the dark matter halo along the line of sight. We then apply our framework to the massive cluster Abell 1689, which features unparalleled multi-wavelength data. In accordance with previous works, we find that the cluster is significantly elongated along the line of sight. Accounting for line-of-sight projections, we require a non-thermal pressure support of $30\text{-}40\%$ at $r_{500}$ to match the gas and weak lensing observables. The joint model retrieves a concentration $c_{200}\sim7$, which is lower and more realistic than the high concentration retrieved from weak lensing data alone under the assumption of spherical symmetry ($c_{200}\sim15$). Application of our method to a larger sample will allow us to study at the same time the shape of dark matter mass profiles and the level of non-thermal pressure support in galaxy clusters.
 
    }


\maketitle

\section{Introduction}

Galaxy clusters stand as the most massive gravitationally bound structures in the Universe. According to the hierarchical structure formation scenario, they develop from primordial overdensities during the early Universe, evolving through mergers and accretion until attaining masses around $10^{14}\text{-}10^{15}M_\odot$ in the present day \citep{Kravtsov_2012}. While galaxy clusters can be observed out to $z\sim2$, in the bottom-up structure formation scenario the most massive clusters are found at low redshifts. Primordial gas is trapped and heated in the gravitational well of the dark matter halo, reaching temperatures on the order of $10^{7}-10^{8}\textrm{K}$. Referred to as the intracluster medium (ICM), this diffuse and hot plasma accounts for the majority ($\sim90\%$) of the baryonic content of galaxy clusters, representing 10-15$\%$ of the total mass. In contrast, the hundreds to thousands of galaxies within the cluster contribute to the stellar mass, which only constitutes 1-2$\%$ of the total mass. Most of the mass content is in the form of dark matter (DM), accounting for the remaining 85$\%$ \citep{Gonzalez_2013,Chiu_2018,Eckert_2022}. 

Tracing the nodes of the cosmic web, galaxy clusters serve as powerful cosmological probes through their mass and redshift distribution \citep{Allen_2011}. The primary systematic uncertainty affecting cosmological inferences with this technique is the mass scale of the detected systems. Thanks to their multi-components structure (ICM, galaxies and DM), galaxy clusters offer multiple observables from which one can constrain the astrophysical processes and recover improved mass measurements \citep{Pratt_2019}. 

In this work, we focus on three of these observables: X-ray emission, Sunyaev-Zel’dovich (SZ) effect, and weak gravitational lensing. Being almost fully ionized, the ICM emits X-rays via thermal bremsstrahlung and line emission. CMB photons crossing the ICM can get inverse-Compton scattered by high-energy electrons of the ICM, resulting in an observable spectral shift known as the Sunyaev-Zel'dovich effect. The shape of ICM thermodynamic properties is set by the total gravitational potential of the system. Dark matter dominating the mass content, one can consider the X-ray emission and SZ signal to be indirect observables of the DM distribution \citep{Bulbul_2010, Capelo_2012,Ettori_2023}. Assuming that the gas pressure balances gravity, the total mass of the cluster can be inferred. A more direct probe of the underlying mass is through gravitational lensing. The deep gravitational potential of a galaxy cluster acts as a lens and changes the path of the light emitted by background galaxies, modifying their apparent shapes and sizes \citep{Bartelmann_2001, Kneib_2011, hoekstra_2013}. This deformation is only dependent on the surface mass density $\Sigma$ of the lens in the plane of the sky and the distances to the lens and the background sources. In this study, we utilize the weak lensing (WL) regime, applicable when the lensing signal is measured at a distance from the center of the lens where the surface mass density is low compared to the critical value $\Sigma_{\rm crit}$ \citep{Umetsu_2020}. In this regime, two main effects are observed: shear ($\gamma$), which increases the ellipticity of the source, and convergence ($\kappa$), which increases the size of the source. As implied by its name, the weak gravitational lensing signal is subtle and requires statistical measurement across multiple sources to be detected reliably.

Assuming hydrostatic equilibrium (HSE), X-ray and SZ observations of well-spatially resolved galaxy clusters can be used to infer a hydrostatic mass profile \citep{pointecouteau_2005, Vikhlinin_2006}. While traditionally the hydrostatic mass was reconstructed using X-ray surface brightness and spectroscopic temperature, in recent years this framework has been extended to model the SZ signal \citep{Ameglio_2007, Ettori_2019, Mu_oz_Echeverr_a_2023, Mastromarino_2024}. This can be achieved by forward modeling a chosen density profile parameterization to the X-ray/SZ data \citep{Eckert_2022a}.  

Reconstructions of mass profiles based on gas properties assume that the gas is in equilibrium within the potential well set by the DM and that the energy is fully thermalized. Galaxy cluster masses determined through this technique show a systematic underestimation when compared to weak lensing measurements: the hydrostatic mass bias \citep{von_der_Linden_2014, Hoekstra_2015, Okabe_2016, Medezinski_2017, Sereno_2017, Herbonnet_2020}. By assuming the gas pressure to be exclusively thermal, the recovered mass will be underestimated if an additional pressure component of non-thermal origin partakes in the total pressure of the ICM. Simulations show that non-thermal pressure support can indeed represent a significant contribution to the total pressure, increasing at large radii where gas tends to be less relaxed \citep{Lau_2009, Meneghetti_2010, Rasia_2012, Angelinelli_2020, Barnes_2021, Bennett_2022}.  

On the other hand, cluster masses estimated using gravitational lensing are independent of the dynamical state and are expected to be close to unbiased on average \citep{Umetsu_2020, Grandis_2021}, with a remaining bias of $<20\%$ that can be calibrated with simulations \citep{EuclidprepGiocoli_2024}. Nonetheless, weak lensing signal is strongly impacted by projection effects. Due to their formation history, galaxy clusters are expected to exhibit triaxial shapes \citep{De_Filippis_2005, Lau_2011, Limousin_2013, Sereno_2018, Lau_2020}, and alignment of their major axis with the line-of-sight will result in a boosted weak lensing signal \citep{Gavazzi_2005,Sereno_2011, Chen_2020, Wu_2022, EuclidprepGiocoli_2024}. Assuming random orientations of the selected galaxy clusters, this orientation bias will be small on average but potentially significant on mass measurements of individual clusters \citep{Despali_2014}. This can be particularly significant of the population of strong lensing clusters for example \citep{Hennawi_2007}.

Given the strengths and weaknesses of the various mass reconstruction techniques, in case multiple observables are available for the same system, the accuracy of the reconstruction can be improved by jointly fitting the observables rather than considering them individually \citep{Sereno_2018}. The different line-of-sight dependencies of the X-ray, SZ, and WL observables allow to constrain the three-dimensional shape of the underlying system in a triaxial geometry \citep{Limousin_2013,Morandi_2012,Sereno_2013,Umetsu_2015,Chiu_2018b}. The comparison between hydrostatic and lensing measurements also provides constraints on deviations from hydrostatic equilibrium \citep{Sayers_2021,Umetsu_2022}. The expected improvement in WL data quality thanks to upcoming wide-field telescopes like \emph{Euclid} and Rubin calls for a substantial effort in the modeling of galaxy cluster properties.

In this paper, we develop a novel model to reconstruct galaxy cluster mass profiles from multi-probe observations (X-ray, WL, SZ). This joint analysis aims at fitting a parameterized density profile to the entire data set. By adding two additional ingredients, line of sight (hereafter LOS) elongation and non-thermal pressure support, we provide a framework that is well suited for the analysis of massive clusters with high data quality. As a test case, we apply our framework to the famous lensing cluster Abell 1689, for which high-quality X-ray, WL and SZ data exist. In Sect.2, we present the details of our modeling for each observable. Sect. 3 introduces Abell 1689 as our case study and the details of the data extraction. Finally, Sect. 4 and Sect. 5 present our results and their interpretation, respectively.  

Throughout this study, we adopt the WMAP9 cosmology \citep{Hinshaw_2013}, with matter density parameter $\Omega_m = 0.2865$, cosmological constant density parameter $\Omega_\Lambda = 0.7134$ and Hubble constant $H_0 = 69.32 \rm{km\ s^{-1}Mpc^{-1}}$. All distances are given in proper units. At the redshift of A1689 (z=0.183), this translates into 1 arcmin = 186.7kpc. The errors displayed in our analysis results are defined as the 1 $\sigma$ errors, i.e. the 16th and 84th percentiles of the posterior parameter distributions.

\section{Mass modeling}

\subsection{Mass profiles}

Galaxy cluster masses are typically computed as the spherically enclosed mass within a chosen radius around the cluster center. This radius is often chosen to match a given level $\Delta$ of over-density compared to the critical density of the Universe, i.e. where $\bar \rho(r_{\Delta}) = \Delta \rho_{\rm crit}$. Common choices for $\Delta$ are 200 or 500.

Using N-body simulations, \citet{Navarro_1996} established the Navarro Frenk White (NFW) parameterization as a good fit to DM halo shapes in the Cold Dark Matter (CDM) paradigm across a wide range of halo masses, from dwarf galaxies to the most massive galaxy clusters \citep{Bullock_2001, Power_2003, Navarro_2004, Macci__2008, Klypin_2016, Le_Brun_2017}. This self-similarity of CDM halos is emphasized by the asymptotic behavior of the inner ($\rho(r) = r^{-1}$) and outer ($\rho(r) = r^{-3}$) slopes of the NFW profile. For a given overdensity $\Delta$, the NFW profile reads

\begin{equation}
    \rho_{\rm NFW}(r) = \frac{\rho_{s}}{\frac{c_\Delta r}{r_{\Delta}} \left(1+ \frac{c_\Delta r}{r_{\Delta}}\right)^2}.
\label{eq:nfw}
\end{equation}

\noindent Here $r_{\Delta}$ is the radius $r$ at which $\bar\rho(r) = \Delta \rho_{\rm crit}$ and $c_\Delta$ is the concentration parameter, $c_\Delta=r_{\Delta} / r_s$ with $r_s$ the scale radius (defined as the radius at which the NFW slope is equal to $-2$). The scaled density $\rho_{s}$ can be reformulated as a function of the critical density of the Universe $\rho_c(z)$ as

\begin{equation}
    \rho_{s} = \frac{\Delta}{3} \frac{c_\Delta^3}{\ln(1+c_\Delta)-\frac{c_\Delta}{1+c_\Delta}}\rho_c(z_l).
\label{eq:rhoc}
\end{equation}

Further studies found that the \citet{einasto1965} profile which describes the density profile as a power law with a rolling index, provides a better representation of DM profiles that stray further from the expected self-similarity in $N$-body simulations \citep{Navarro_2004, Klypin_2016, Ludlow_2016, Brown_2020}:

\begin{equation}
    \rho_{\rm Einasto}(r) = \rho_s \exp{\left[-\frac{2}{\alpha}\left(\left( \frac{r}{r_s}\right)^{\alpha} -1 \right)\right]},
\label{eq:einasto}
\end{equation}

\noindent with $\alpha$ the shape parameter responsible of the curvature of the profile. The CDM paradigm makes strong predictions for the value $\alpha$. \citet{Navarro_2009} and \citet{ Ludlow_2016} showed that CDM halos can be well described by an Einasto profile with $\alpha\sim0.18$, which is further supported by observations \citep{Umetsu_2014, Eckert_2022}.

\subsection{Observables}
The different components (ICM, galaxies, DM) of galaxy clusters provide for a wide range of observables. In this work, we build a global framework to reconstruct the mass distribution within galaxy clusters by jointly fitting the gas observables - the X-ray emission and the Sunyaev-Zel'dovich effect - and the weak gravitational lensing signal in a self-consistent way.

\subsubsection{X-ray observables}

Being heated to high temperatures, the ICM emits X-rays via bremsstrahlung and line emission and allows us to recover two primary observables: X-ray surface brightness and spectroscopic temperature.
The observed surface brightness is the integrated emissivity of all the gas particles in a given line of sight. Each volume element will thus have a contribution $\Lambda \left( T, Z \right)n_e n_H$ with $n_{e}$ the electronic density, $n_H$ the proton density, and $\Lambda \left( T, Z \right)$ the cooling function. In a soft X-ray energy band (e.g. 0.5-2 keV), $\Lambda(T,Z)$ is roughly constant for temperatures greater than $\sim2$ keV. The surface brightness is then given by the gas emissivity integrated along the LOS,
\begin{equation}
    \begin{split}
    S_X(r) = \int \Lambda \left( T, Z \right)n_e n_H \,\mathrm{d}l \\ \propto \int n_e^2\,\mathrm{d}l.
    \end{split}
    \label{eq:sx}
\end{equation}

The ratio of $n_{e}$ to $n_{H}$ is determined by the level of ionization of the ICM and is assumed to be constant throughout the cluster's volume. In a fully ionized plasma $n_e/n_H\sim1.17$ assuming Solar abundance ratios as given in \citet{Asplund2009}.

Similarly, the measured spectroscopic temperatures depend on the projection of the 3D temperatures of volume elements along the LOS:
\begin{equation}
    T_X = \frac{\int w T_{3D} \,\mathrm{d}l}{\int w \,\mathrm{d}l} = \frac{\int n_e^2 T_{3D}^{-3/4} T_{3D} \,\mathrm{d}l}{\int n_e^2 T_{3D}^{-3/4} \,\mathrm{d}l}.
\label{eq:Tx}
\end{equation}

As shown in Eq. \ref{eq:Tx}, the projected temperature depends on the emissivity-dependent weight $w$ given to each volume element, which at high temperatures ($T>3$ keV) can be written as $w = n_e^2 T_{3D}^{-3/4}$ \citep{Mazzotta_2004}.

\subsubsection{The Sunyaev-Zel'dovich effect}

The SZ effect is the inverse Compton scattering of CMB photons with the electrons of the hot ICM. The amplitude of the resulting spectral distortion is proportional to the Compton parameter $y$ and is directly related to the energy $k_B T_{3D}$ of the encountered charged particles and their density $n_e$ (see Eq. \ref{eq:sz}). As a result, the projected SZ signal is proportional to the gas pressure integrated along the LOS,

\begin{equation}
    y = \int \frac{k_{B}T_{3D}}{m_{e}c^2}n_{e}\sigma_{T} \, \mathrm{d}l ,
\label{eq:sz}
\end{equation}

with $\sigma_T$ the Thompson cross-section.

\subsubsection{Weak gravitational lensing}

The lensing signal depends on the projected mass density along the LOS, the surface mass density $\Sigma$:
\begin{equation}
       \Sigma(\theta) = 2 \int_{\theta}^{\infty} \frac{\rho(\theta')\theta'}{\sqrt{\theta'^{2}-\theta^{2}}}\mathrm{d}\theta' ,
\label{eq:sigma}
\end{equation}
where $\theta$ is the angular separation from the center of the lens.

The foreground galaxy cluster acts like a lens located between the observer and the background galaxies. Similar to the case of an optical lens, the observed signal depends on the angular diameter distances $D_L$ (distance from the observer to the lens), $D_S$ (distance from the observer to the source), and $D_{LS}$ (distance between the lens and the source). These quantities are encapsulated in the critical surface density 
\begin{equation}
    \Sigma_{\rm crit} = \frac{c^2}{4 \pi G} \frac{D_S}{D_{LS}D_L}.
\label{eq:sigcrit}
\end{equation}

The weak lensing regime is defined as the regime where $\Sigma \ll \Sigma_{\rm crit}$. In this regime, the lensing  effect is not detectable on individual galaxies and the signal can be brought up by averaging over multiple background sources.

We briefly summarize here the main equations of cluster weak lensing. For more details we refer the reader to \citet{Umetsu_2020}. In a polar coordinate system centered on the lens, the shear $\gamma$ can be split into two geometrically defined components, the tangential shear $\gamma_+$, and the 45° rotated cross shear $\gamma_{\times}$ :

\begin{equation}
\begin{split}
\gamma_+(\theta, \phi) = - \gamma_1(\theta)\cos{2\phi} - \gamma_2(\theta)\sin{2\phi}, \\ \gamma_\times(\theta, \phi) = + \gamma_1(\theta)\sin{2\phi} - \gamma_2(\theta)\cos{2\phi},    
\end{split}
\end{equation}

with $\gamma_1$ and $ \gamma_2$ the spin-2 shear components in Cartesian coordinates and ($\theta$, $\phi$) the polar coordinates (distance to the center and angle, respectively). 

This formulation has the advantage of providing a consistent sanity check: under the approximation of weak lensing, $\left| \gamma(\theta) \right| \ll 1$, one finds 

\begin{equation}
\begin{split}
\gamma_+(\theta) = \frac{\Delta \Sigma(\theta)}{\Sigma_{\rm crit}} , \\ \gamma_\times(\theta) = 0 .   
\end{split}
\end{equation}
And thus, in the absence of systematics, the 45° rotated cross shear $\gamma_{\times}$ must be consistent with zero. Here we have introduced the excess surface mass density $\Delta \Sigma(\theta)$, which is defined as the difference between the average surface mass density $\Bar{\Sigma}(\theta)$ within a circle of radius $\theta$ and the local surface mass density,

\begin{equation}
\begin{split}
   \Delta \Sigma(\theta) = \Bar{\Sigma}(\theta) - \Sigma({\theta}), \\
   \Bar{\Sigma}(\theta) = \frac{2}{\theta^2} \int_{0}^{\theta} \theta' \Sigma(\theta')\mathrm{d}\theta'.
\label{eq:dsigma}
\end{split}
\end{equation}

The averaged ellipticity of the background galaxies, the reduced tangential shear $g_+ (\theta)$, offers a direct observable of the weak lensing effect. It is related to the shear as

\begin{equation}
    g_+(\theta) = \frac{\gamma_+(\theta)}{1 - \kappa(\theta)}.
\end{equation}

\noindent with $\kappa=\Sigma/\Sigma_{\rm crit}$ the convergence. 





In the linear regime:

\begin{equation}
    g_+(\theta) \simeq \gamma_+(\theta).
    \label{eq:shear_approx}
\end{equation}

In this work, to probe high surface mass density regions, thus considering the non-linear but subcritical regime, we will use the following formulation of the mean tangential shear \citep{seitz1996steps}:

\begin{equation}
\langle g_+(\theta) \rangle = \frac{\langle \gamma_+(\theta) \rangle}{1 - f_l\langle \kappa(\theta) \rangle} ,
\label{eq:shear1996}
\end{equation}

\noindent with 

\begin{equation}
    f_l = \frac{\langle \Sigma_{\text{\rm crit}}^{-2} \rangle}{\langle \Sigma_{\text{\rm crit}}^{-1} \rangle ^2},
\end{equation}

\noindent where the brackets denote the averaging over the source redshift distribution. 

\subsection{Modeling}
\label{sec:mass_modeling}

\begin{table*}
\caption{Prior distributions on the model parameters.}              
\label{table:priors}      
\centering                                      
\begin{tabular}{l l l l}          
\hline\hline                        
 & Parameter & \multicolumn{2}{l}{Prior} \\    
\hline                                   
NFW & $c_{200}$ & \multicolumn{2}{l}{$\mathcal{U}([1, 20])$} \\  
 & $r_{200}$ [kpc] & \multicolumn{2}{l}{$\mathcal{U}([900, 4000])$} \\  
\hline
Einasto & $c_{\text{norm}}$ & \multicolumn{2}{l}{$\mathcal{N}(\mu=1.8, \sigma=1.5, \text{min}=0, \text{max}=5)$} \\  
 & $r_s$ [kpc] & \multicolumn{2}{l}{$\mathcal{N}(\mu=700, \sigma=300, \text{min}=100, \text{max}=1800)$} \\  
 & $\mu$ & \multicolumn{2}{l}{$\mathcal{N}(\mu=5, \sigma=3, \text{min}=1, \text{max}=20)$} \\  
\hline
$P_{NT}^{polytropic}$ & $\log_{10}(P_{0\text{NT}}[\text{keV.cm}^3])$ & \multicolumn{2}{l}{$\mathcal{U}([-5, -2])$} \\  
 & $\beta$ & \multicolumn{2}{l}{$\mathcal{N}(\mu=0.9, \sigma=0.13)$} \\  
\hline


$P_{NT}^{Angelinelli}$ & $a_0$ & \multicolumn{2}{l}{$\mathcal{U}([-0.5, 2])$} \\  
 & $a_1$ & \multicolumn{2}{l}{$\mathcal{U}([-1, 2])$} \\  
 & $\log_{10}(a_2)$ & \multicolumn{2}{l}{$\mathcal{U}([-6, 2])$} \\  
\hline

\hline
 & Elongation $e$ & \multicolumn{2}{l}{$\mathcal{N}(\mu=1.0, \sigma=0.2, \text{min}=0.1, \text{max}=10)$} \\  
 & External pressure $P_0$ & \multicolumn{2}{l}{$\mathcal{N}(\mu = P_{SZ}(r_{\text{max}}), \sigma = \text{err}(P_{SZ}(r_{\text{max}})))$} \\  
\hline                                             
\end{tabular}
\end{table*}

This work builds upon the Python package \begin{tt}hydromass\end{tt}\footnote{\url{https://github.com/domeckert/hydromass}} \citep{Eckert_2022a}, which allows to model X-ray and SZ observables jointly under the assumption that the gas is in hydrostatic equilibrium within the potential well set by the total mass distribution. Here we extend the framework to include a self-consistent modeling of WL signal and gas observables. This hybrid SZ/X-ray method uses a Bayesian forward modeling approach in which 3D quantities are sampled according to priors and then projected along the LOS to be compared with the data. The optimization is done using the No U-Turn Sampler \citep{hoffman_2011} implemented within the \begin{tt}PyMC\end{tt} package \citep{Patil_2010}, running simultaneously 4 chains of 4000 steps each with an equal number of tuning steps.

As introduced in \citet{Eckert_2020}, \begin{tt}hydromass\end{tt} uses a linear combination of King functions $\Phi_k(r)$ to model the 3D emissivity profile $\epsilon(r)$:

\begin{equation}
    \epsilon(r)=\sum_{k=1}^{N_{\mathrm{K}}} \alpha_k \Phi_k(r),
\end{equation}

\noindent with $\alpha_k$ the coefficient of the basis function $\Phi_k(r)$. The basis functions are chosen to be flexible enough to reproduce a wide range of profiles, under the assumption that the underlying 3D profile is smooth, positive definite and monotonically decreasing. 

To infer the shape of the mass profile, two main methods are implemented within the \begin{tt}hydromass\end{tt} package \citep{Ettori_2011}:\\
\begin{itemize}

     \item Mass model: a functional form for the mass profile (e.g. NFW, see Eq. \ref{eq:nfw}) is assumed and the model optimizes the parameters of the mass profile to reproduce the data at hand.

    Assuming hydrostatic equilibrium (Eq. \ref{eq:hse}), the 3D pressure can be recovered by integrating the modeled mass profile $M_{\mathrm{mod}}$ (Eq. \ref{eq:p3d}),

    \begin{equation}
    \frac{\mathrm{d} P_{\mathrm{gas}}}{\mathrm{d} r}=-\rho_{\mathrm{gas}} \frac{G M_{\mathrm{tot}}(<r)}{r^2},
    \label{eq:hse}
    \end{equation}
    
    \begin{equation}
    P_{3 \mathrm{D}}(r)=P_0+\int_r^{r_0} \frac{\rho_{\mathrm{gas}} G M_{\mathrm{mod}}\left(r^{\prime}\right)}{r^2} \mathrm{~d} r^{\prime}.
    \label{eq:p3d}
    \end{equation}
    
    \item Parametric forward: the pressure profile is modeled as a generalized NFW (gNFW) \citep{Nagai_2007} :
    \begin{equation}
    P_{\text {forw }}(r)=\frac{P_0}{\left(r / r_{\mathrm{s}}\right)^c\left(1+\left(r / r_{\mathrm{s}}\right)^a\right)^{\frac{b-\gamma}{a}}}.
    \label{eq:gnfw_pressure}
    \end{equation}
     The pressure profile is projected along the LOS assuming spherical symmetry and fitted to the data. The mass profile is then inferred from the best fitting 3D pressure profile.
   
\end{itemize}

This work focuses on the former method (mass model) by implementing an option to jointly fit the WL data with the X-ray and SZ information. To accommodate potential deviations from hydrostatic equilibrium and spherical symmetry, we introduce two additional parameters to the model: non-thermal pressure and LOS elongation. The parametric forward method is used as a reference to validate our model.

\subsubsection{From density profile to mean tangential shear}

Our implementation of WL in the \begin{tt}hydromass\end{tt} package is fully numerical and it can thus include any parametric mass model. Here we focus on the NFW and Einasto models, which are most commonly used to describe DM halos. Following Eq. \ref{eq:nfw}, the free parameters of the NFW density profile are the concentration $c_\Delta$ and the overdensity radius $r_{\Delta}$ from which the enclosed mass can easily be recovered:
\begin{equation}
    M_{\Delta} = \frac{4 \pi \Delta}{3} \rho_{\rm crit} r_{\Delta}^{3}.
    \label{eq:mdelta}
\end{equation}
We make the choice of $\Delta = 200$ which is the most common definition of the NFW concentration 
\citep{Wechsler_2002, Klypin_2016}.

The Einasto profile (Eq. \ref{eq:einasto}) is implemented according to the optimized parameterization presented by \citet{Ettori_2019}. The fitted parameters are the scale radius $r_s$, the inverse shape parameter $\mu = 1/\alpha$ and the unitless normalization parameter $c_{\rm norm}$ such that :
\begin{equation}
    \rho_s = \frac{\Delta \rho_{\rm crit}}{3}\left[\ln(1+c_{\rm norm})-\frac{c_{\rm norm}}{1+c_{\rm norm}}\right].
    \label{eq:einasto_c}
\end{equation}
The priors on the model parameters for both mass models can be found in Table \ref{table:priors}.

The first step in the numerical modeling of the mean tangential shear is to obtain the surface mass density $\Sigma$ by projecting the density profile (see Eq. \ref{eq:sigma}). The surface mass density is then integrated in circular bins to compute $\Delta \Sigma$ according to Eq. \ref{eq:dsigma}. We use Eq. \ref{eq:shear1996} to model the mean tangential shear, allowing us to use the non-linear but subcritical regime to probe the density profile of the cluster at lower radii than the traditional linear weak lensing approximation (Eq. \ref{eq:shear_approx}).

\subsubsection{Non-thermal pressure}
\label{sec:model_NTpressure}

Since the modeling of the gas properties assumes hydrostatic equilibrium while the WL signal does not depend on dynamical state, following \citet{Eckert_2019} we modify the hydrostatic equilibrium (Eq. \ref{eq:hse}) to include a non-thermal correction term,
\begin{equation}
    \frac{\text{d}}{\text{d}r} \left( P_{\text{th}}(r) + P_{\text{NT}}(r) \right) = -\rho_{\text{gas}} \frac{GM_{\text{tot}}(<r)}{r^2},
\label{eq:pnt}
\end{equation}
with $P_{\text{th}}$ and $P_{\text{NT}}$ the thermal and non-thermal pressure, respectively. The non-thermal pressure term therefore represents any additional pressure term that is required to balance the local gravitational field. This term is expected to be dominated by merger-induced random gas motions that have not yet thermalized \citep{Lau_2009, Nelson_2014, Biffi_2016, Vazza_2018}, although any additional non-thermal contribution from, e.g., magnetic fields or cosmic rays, is included in the non-thermal pressure term. Throughout this work, we characterize the relative importance of non-thermal pressure using the non-thermal pressure fraction,
\begin{equation}
    \alpha_{\operatorname{NT}}(r) = \frac{P_{\text{NT}}(r)}{P_{\text{tot}}(r)}.
    \label{eq:alpha}
\end{equation}
First-order development shows that $\alpha_{NT}\approx1-M_{\rm HSE}/M_{\rm tot}$ \citep{Eckert_2019,Ettori_2021}. We describe the non-thermal pressure profile using a parametric functional form and optimize for the corresponding model parameters. We test two different parameterizations of the non-thermal pressure to assess the dependence of the recovered parameters on the chosen parametric form.
\begin{itemize}
    \item Using a set of 20 clusters simulated at high resolution with the adaptive mesh refinement code ENZO, \citet{Angelinelli_2020} showed that non-thermal pressure profiles in simulated clusters can be described as: 
    \begin{equation}
        \alpha_{\operatorname{NT}}^{\rm Angelinelli}(r) = a_0 \cdot \left(\frac{r}{r_{200, m}}\right)^{a_1} + a_2,
    \label{eq:pnt_angelinelli}
    \end{equation}
    \noindent with $r_{200,m}$ the radius within which the mean density is 200 times the mean matter density of the Universe (as opposed to the critical density). This parameterization describes the radial profile of $\alpha_{\operatorname{NT}}$ as a power law (described by the parameters $a_0$ and $a_1$) with a central floor ($a_2$).  

    \item \citet{Ettori_2021} suggested a different parameterization of the non-thermal pressure, in which the non-thermal pressure is assumed to scale with the electron density $n_e$ similarly to the thermal pressure,
    \begin{equation}
        P_{\text{NT}}^{\text{Polytropic}}(r) = P_{0\text{NT}} \left(\frac{n_e(r)}{n_0}\right)^\beta.
        \label{eq:pnt_ettori}
    \end{equation}
    Here the index $\beta$ governs the shape of the non-thermal pressure profile, $n_0=10^{-3}$ cm$^{-3}$ is a scale density, and $P_{0\text{NT}}$ describes the non-thermal pressure at $n_e=n_0$. This functional form is motivated by the expected dependence of the various sources of non-thermal pressure (turbulence, magnetic fields, cosmic rays) on the gas density \citep{Miniati_2015}. The index $\beta$ is expected to be around $\sim0.9$ \citep{Ettori_2021}, which is less than the polytropic index of the thermal pressure \citep[$\Gamma\sim1.2$,][]{Capelo_2012,Ghirardini_2019}, such that generally $\alpha_{\operatorname{NT}}$ increases with radius.
\end{itemize}

To set up the priors on the parameters of the non-thermal pressure profiles, we used the set of 20 simulated clusters presented in \citet{Angelinelli_2020} and fitted their non-thermal pressure profile with Eq. \ref{eq:pnt_ettori}. We then used the distributions of the model parameters in the simulated sample to set up the priors on the parameters when fitting the data. The priors are broad enough to encompass the range of profiles in the simulated data set. Our implementation of the polytropic model uses a log uniform prior on $P_{0 N T}$ and a Gaussian prior on $\beta$, and uniform priors for the Angelinelli model to remain as uninformative as possible. The priors on the parameters of both models are shown in Table \ref{table:priors}.

\subsubsection{Elongation}

The model implemented here accounts for deviations from spherical symmetry by implementing a correction factor $e$ that describes the ratio between the true projected quantity and the quantity calculated assuming spherical symmetry. While methods to reconstruct the shape parameters of DM and gas in a triaxial geometry have been developed \citep[e.g.][]{Sereno_2017_CL3D,chexmate_2023}, these methods are computationally expensive and still make the hypothesis of a single ellipsoidal halo. Here we attempt to account for the impact of deviations from spherical symmetry using a simpler one-dimensional correction factor.

We consider the 3D distribution $X(x,y,z)$ of a generic quantity $X$  (X-ray emissivity, thermal pressure, mass density) and the corresponding quantity $X_p(x,y)$ projected along axis $z$ (X-ray surface brightness $S_X$, SZ Compton parameter $y$, or surface mass density $\Sigma$). The projected quantity is given by

\begin{equation}
    X_p(x,y) = \int_{-\infty}^\infty X(x,y,z)\,\mbox{d}z.
\end{equation}

The observed quantity is the circularly averaged projected profile $\bar X_p(\theta_p)$, with $\theta_p$ the projected angular radius. The discretized profile $\{X_{p,i}\}_{i=1}^N$ is given by the average of $X_p$ in the circular annulus enclosed within projected angular radii $\theta_{p,i}$ and $\theta_{p,i+1}$, with $\{\theta_{p,i}\}_{i=1}^N$ the boundaries of the defined annular bins. The measured values can therefore be written as
\begin{equation}
X_{p,i} = \frac{1}{\pi(\theta_{p,i+1}^2 - \theta_{p,i}^2)}\int X_p(x,y)\, \mbox{d}\Omega,
\end{equation}
where $x$ and $y$ are the angular coordinates and $\Omega$ the corresponding solid angle.

Under spherical symmetry, this quantity can be estimated by convolving the mean discretized 3D profile $\{X_i\}_{i=1}^N$ with the projected volumes:
\begin{equation}
    X_{p,i} = \sum_{j=1}^N V_{s;i,j} X_j,
    \label{eq:proj_vol}
\end{equation}
\noindent where the projection volumes $V_{s;i,j}$ are given by the intersection of spherical shell $j$ with cylindrical shell $i$ \citep{Kriss_1983}. In the general case where the system is not spherical, we define the elongation factor $e$ as the mean ratio between the (a priori unknown) true 3D volumes $V_{t}$ and their spherical counterparts $V_{s}$ \citep{De_Filippis_2005},
\begin{equation}
    e = \left\langle \frac{V_{t}}{V_s} \right\rangle.
    \label{eq:e_def}
\end{equation}

\begin{figure}
\resizebox{\hsize}{!}{\includegraphics[width=\textwidth]{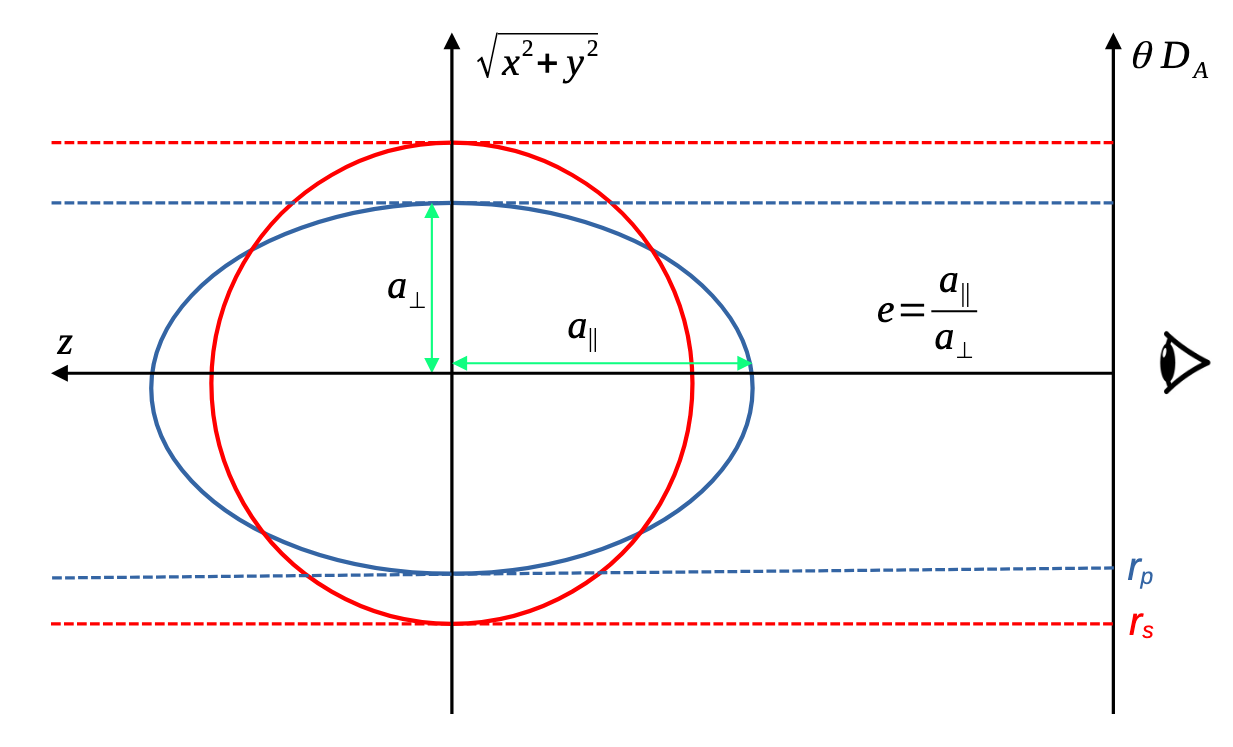}}
\caption{\label{fig:elong_drawing}Sketch of the implemented elongation correction. The sketch shows the isodensity surface of an ellipsoidal halo (blue solid curve) in the LOS ($z$) - plane of the sky ($\sqrt{x^2+y^2}$), compared with a spherical halo with the same 3D radially-averaged profile (red solid line). The projected profile (dashed lines) is boosted by a factor $e=a_{\parallel}/a_{\perp}$ and compressed on the plane of the sky by a factor $r_s/r_p=e^{1/3}$ (Eq. \ref{eq:radius_stretch} and \ref{eq:elongprior}).}
\label{fig:elong_drawing}
\end{figure}

The meaning of the elongation factor is described in Fig. \ref{fig:elong_drawing} for the case of a triaxial halo that is perfectly aligned with the LOS. Let $\{X_{s,i}\}_{i=1}^N$ be the projection of $\{X_i\}_{i=1}^N$ assuming spherical symmetry, i.e. the profile of a spherical halo with the same radial 3D profile as the actual halo (red solid circle in Fig. \ref{fig:elong_drawing}). The figure shows that the projected profile is boosted or suppressed by a factor $e$ with respect to the spherical case, as the LOS volumes are altered. On top of that, the radial axis is stretched or compressed by a factor $e^{1/3}$, such that the projected plane-of-the-sky radius is related to the spherically averaged radius as 
\begin{equation}
    r_{s,i} = D_A \theta_{p,i} e^{1/3} = r_{p,i}e^{1/3}.
\label{eq:radius_stretch}
\end{equation}
Therefore, the measured radial profile is related to the spherically averaged projected profile $X_s$ as \citep{Sereno_2007}
\begin{equation}
X_{p}(r_p) = X_s(r_s) \times e .\label{eq:elong_vs_spherical}
\end{equation}
As a result, every projected quantity gets modified by a factor $e$ and stretched by a factor $e^{1/3}$ with respect to the spherical case, such that the measured projected profiles become
\begin{eqnarray}
y(r_p) & = & y_s(r_s) \times e , \label{eq:elong_y} \\
S_X(r_p) & = & S_{X,s}(r_s) \times e ,\\
\langle g_+(r_p) \rangle & \simeq & \langle g_{+,s}(r_s) \rangle  \times e ,
    \label{eq:elongprior}
\end{eqnarray}
where the quantities with the index $s$ denote the spherically averaged profiles projected under spherical symmetry (Eq. \ref{eq:proj_vol}). We make the fair assumption that the observed temperature $T_X$ should not be impacted by the LOS elongation, as the elongation factors at the numerator and denominator approximately cancel out (see Eq. \ref{eq:Tx}). 

At every step of our optimization process, we solve Eqs. \ref{eq:elong_y} through \ref{eq:elongprior} to estimate the model profile at every observed data point. Numerically solving this equation requires applying a root-finding algorithm at every point, which is computationally expensive. In a realistic case $e\approx1$, and $|r_s-r_p|/r_s \ll 1$, such that Eq. \ref{eq:elong_vs_spherical} can be linearly expanded:
\begin{eqnarray}
    X_p(r_p) & \approx & e \left( X_s(r_p) + \frac{\partial X_s}{\partial r} (r_p) (r_s - r_p) \right), \\
    \, & \approx & e X_s(r_p) \left( 1 + \frac{\partial \ln X_s}{\partial \ln r} (r_p) (e^{1/3} - 1) \right). \label{eq:elong_corrfact}
\end{eqnarray}
In realistic cases, we find that the elongation-dependent correction factor in the right-hand side of Eq. \ref{eq:elong_corrfact} reproduces the true solution of Eq. \ref{eq:elong_vs_spherical} with an accuracy of a few percent at every radius.

In practice, the constraints on the elongation parameter $e$ are obtained from the ratio of the pressure inferred from X-ray and SZ observations. These observations trace the exact same quantity, albeit with different LOS dependencies (see Eqs. \ref{eq:sx}, \ref{eq:Tx}, \ref{eq:sz}). Since $n_e\propto S_X^{1/2}$, the ratio of spherically deprojected SZ  to X-ray pressure becomes \citep[e.g.][]{Kozmanyan_2019,Ettori_2020},
\begin{equation}
    \frac{P_{SZ}}{P_X} = e^{1/2}.
\end{equation}
In a spherical halo $e=1$ for every LOS. Conversely, in a generic case $e$ is larger than 1 when the system is oriented preferentially along the LOS and smaller than 1 when the system is elongated in the plane of the sky. Therefore, the profiles corrected for elongation (Eq. \ref{eq:elong_vs_spherical}) correspond to the profiles one would observe along a fiducial sight line where $P_{SZ}=P_X$, i.e. to the spherically averaged profiles.

The Gaussian prior used on the elongation parameter can be found in Table \ref{table:priors}. Being centered on 1 (i.e. the spherical case), it assumes that halos are randomly orientated with respect to the LOS. \citet{Lau_2020} used massive halos extracted from large N-body simulations to estimate the distribution of triaxial galaxy clusters shapes. The authors found that deviations from spherical symmetry lead to variations of surface mass density of $\sim20\%$, which we use as the width of our Gaussian prior on $e$. Since we model the mean tangential shear profile from the projected mass density, the way the elongation parameter $e$ propagates (Eq. \ref{eq:elongprior}) implies that we assume a common dark matter and gas geometry. The model also assumes that the impact of asphericity is constant with radius. Future versions of our code will implement a self-consistent calculation of DM and gravitational potential distributions.


\section{Application to A1689: data set}

\begin{figure*}[htbp]
    \centering
    \begin{subfigure}[b]{\linewidth}
      \centering
      \includegraphics[width=\linewidth, trim={0cm 0cm 0cm 0cm},clip]{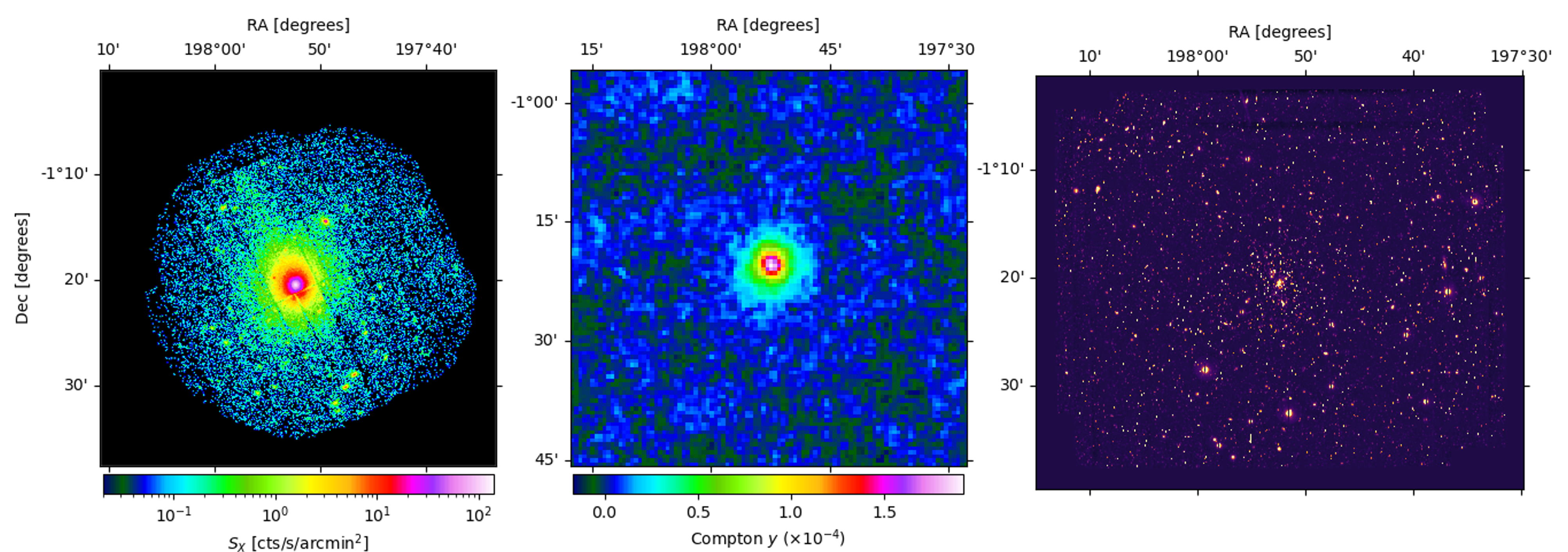}
      \caption{Multi-wavelengths maps of A1689. Left: \emph{XMM-Newton} count map in the [0.7-1.2] keV. Middle: \emph{Planck} + ACT Compton-$y$ map. Right: Subaru Suprime-Cam image ($V$ band).}
      \label{fig:data:images}
    \end{subfigure}
    \hfill
    \begin{subfigure}[b]{\linewidth}
      \centering
      \includegraphics[width=0.8\linewidth, trim={0cm 0cm 0cm 0cm},clip]{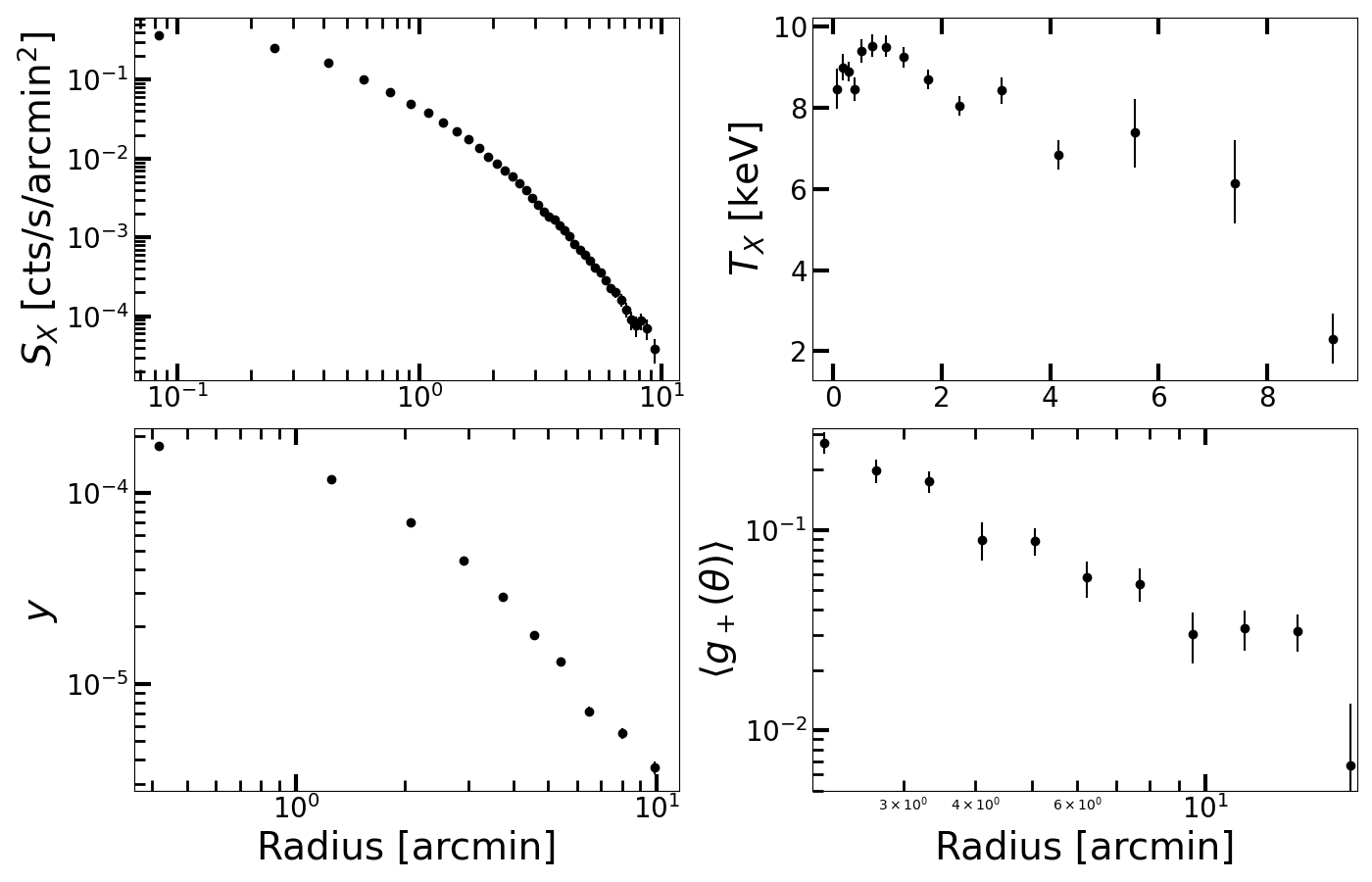}
      \caption{A1689 radial profiles:\\
    upper left: X-ray surface brightness (\emph{XMM-Newton}); upper right: X-ray spectroscopic temperatures (\emph{XMM Newton}); lower left: SZ Compton $y$ (\emph{Planck} + ACT); lower right: Mean tangential shear (Subaru Suprime-Cam and CFHT MegaPrime/MegaCam).
    All of these 1D profiles will be used as inputs separately for single-probe analysis, or together for our joint multi-probe analysis.}
      \label{fig:data:profiles}
    \end{subfigure}
    \caption{A1689 X-ray, SZ and WL data set: 2D images and 1D profiles.}
    
    \label{fig:data}
\end{figure*}

Abell 1689 is a galaxy cluster situated in the Virgo constellation at redshift z = $0.183$. It is one of the most massive clusters in the observable Universe, with a mass usually found above $10^{15}M_{\odot}$. Thanks to its very high mass, A1689 acts as a very powerful gravitational lens. Comparison between different observables of A1689 hint at an unrelaxed and/or asymmetrical situation: the X-ray analysis shows lower mass than the lensing signal \citep{Peng_2009}. Moreover, NFW fits of A1689 show very high concentration values, in tension with the concentration of $\sim4$ predicted by N-body simulations for halo masses around $10^{15}M_{\odot}$. Analyses of the mass profile of A1689 assuming spherical symmetry found much higher concentrations. From a weak lensing only analysis based on the convergence map, \citet{Umetsu_2008} measured $c_{200} = 10.7^{+4.5}_{-2.7}$ and $M_{200}=1.76^{+0.20}_{-0.20}\times 10^{15}M_\odot$. Previous triaxial analyses of the multi-wavelength properties of A1689 showed the system is probably highly elongated along the LOS \citep{Oguri_2005,Morandi_2011,Sereno_2011,Sereno_2012,Umetsu_2015}, which may explain the high concentration inferred from weak lensing studies. 

In addition, Abell 1689 show the presence of a radio halo \citep{Vacca_2011} and in the statical analysis of the radio halo-cluster merger connection by \citet{Cuciti_2015} it was found to be the only radio halo cluster to occupy the region of relaxed clusters in the diagram of “morphological parameters” (c,w). This was explained as due to the fact that this cluster is undergoing a merger event at a very small angle with the line of sight \citep{Andersson_2004}.
\citet{Rossetti_2010} also classified this object as a merger cool core remnant.

Therefore, A1689 is an ideal candidate to evaluate the efficiency of the elongation parameter and non-thermal pressure support we introduced in the model we implemented in \begin{tt}hydromass\end{tt}.

\subsection{X-rays}

The \emph{XMM-Newton} temperature and brightness profiles were extracted using the analysis pipeline described in \citet{bartalucci23} and \citet{rossetti24}, which together serve as the reference data reduction method adopted by the Cluster HEritage project with XMM-Newton: Mass Assembly and Thermodynamics at the Endpoint of Structure Formation collaboration \citep[CHEX-MATE;][]{chexmate_2021}. The details of the procedures for producing the surface brightness and temperature profiles can be found in \citet{bartalucci23} and \citet{rossetti24}, respectively. Here, we briefly summarize the main steps.

We reduced the \emph{XMM-Newton} observation of A1689 (observation ID 0093030101) using the {\tt XMMSAS} software version $16.1$ and applying the latest calibration files\footnote{\url{https://www.cosmos.esa.int/web/xmm-newton/current-calibration-files}}. The data were reprocessed with the {\tt emchain} and {\tt epchain} tools. We performed standard flag and pattern cleaning and we used the {\tt mos/pn-filter} task to remove time periods affected by flaring background. The resulting clean exposure time is $34.6$, $35.3$, and $25.6$ ks for the MOS1, MOS2, and pn camera, respectively. We detected point-sources in two energy bands ($[0.5-2]$ keV and $[2-7]$ keV) and masked them in our analysis \citep[see][for more details]{bartalucci23}.
We extracted images in the $[0.7-1.2]$ keV energy range, with the corresponding exposure and background maps, with the XMMSAS tools {\tt mos/pn-spectra} and {\tt mos/pn-back}. We then derived the surface brightness profile in concentric annuli, centered on the peak of the X-ray emission at $R.A.=197.8725$ and Dec. $-1.3414$ using the Python package {\tt pyproffit} \citep{Eckert_2020}. To remove the emissivity bias associated with the presence of gas density inhomogeneities \citep{Nagai_2011}, we produced a two-dimensional surface brightness map with the Voronoi tessellation technique \citep{Cappellari_2003} and calculated the azimuthal median of the surface brightness within concentric annuli \citep{Eckert_2015}, which allows to mitigate the bias associated to regions of enhanced X-ray emission. The resulting surface brightness profile is shown in the top-left panel of Fig. \ref{fig:data}b.

To derive the temperature profile we extracted spectra in 15 concentric annuli around the X-ray peak, ensuring a roughly constant signal-to-noise ratio in each radial bin. In each region and for each camera, we extracted a spectrum from the observation and the corresponding response matrix (RMF) and ancillary file (ARF) as well as a spectrum of the particle background component, with the tools {\tt mos/pn-spectra} and {\tt mos/pn-back}. With the same tools, we also extract a spectrum from an external annulus between $10.6^\prime$ and the edge of the field-of-view, that we use to estimate the sky background components. As described in detail in \citet{rossetti24}, we build a model of all the background components (quiescent particle background, cosmic X-ray background, Galactic halo, local hot bubble, and residual soft protons) for each region. We then jointly fit the MOS1, MOS2, and pn spectra using the APEC model \citep{APEC} in a Bayesian MCMC framework, treating the parameters of the background model as priors to derive the temperature, metal abundance and normalization of the ICM. The temperature profile is shown in the top-right panel of Fig. \ref{fig:data:profiles}. In each radial bin, the central value is derived as the mean of the posterior distribution of temperature values, while the error bars represent the 68\% confidence interval.

\subsection{Sunyaev-Zel'dovich effect}

The Sunyaev–Zel'dovich Compton $y$ parameter map used in this work combines data from the Atacama Cosmology Telescope (ACT) and $Planck$ with 1.6 arcmin full-width half-maximum beam size, similar to the data presented in \citet{chexmate_2023}. Specifically, the component-separated maps\footnote{\url{https://lambda.gsfc.nasa.gov/product/act/actadv_prod_table.html}} from the ACT data release 6 (DR6) observed at three frequency bands (90, 150, and 220 GHz) were utilized, providing a lower rms noise compared to the DR4 data (see \citealt{Madhavacheril_2020, coulton2023}).

The  Compton $y$ parameter 1D profile shown in Fig. \ref{fig:data:profiles} was extracted by integrating concentric circular bins equidistant in log space up to a maximum aperture of 10 arc minutes. To estimate the noise level of the Compton-$y$ map, equal-sized patches of the ACT maps were randomly sampled. A computed uniform background of $4.3 \times 10^{-6}$ was used to infer the uncertainties of the data points in the 1D profile presented. 
Computing the pressure profile in the mass model framework (Eq. \ref{eq:p3d}) requires a prior on the external pressure $P_0$. We defined the latter as a normal centered on the pressure at the maximal radius aperture radius (e.i. 10 arc minutes) as shown in Table \ref{table:priors}. This value and its error (which is used as the standard deviation of the normal prior) are taken from the 3D $Planck$ pressure profile, part of the official products of the CHEX-MATE consortium.

\subsection{Weak lensing}
\label{sec:wl}

We performed a weak-lensing analysis of A1689 using archival observations from the Subaru Suprime-Cam and CFHT MegaPrime/MegaCam instruments. These data were produced as part of the CHEX-MATE collaboration \citep[][]{chexmate_2021}. Full details of image reduction, photometry, and shape analysis for the AMALGAM2 sample, comprising 40 CHEX-MATE clusters, will be presented in forthcoming publications (Gavazzi et al.; Umetsu et al., in preparation). In this work, we present a concise summary of our weak-lensing analysis of A1689. Our study is based on wide-field imaging in the $B$, $V$, $R_\mathrm{C}$, $i^+$, and $z^+$ bands, obtained with Suprime-Cam on the 8.2~m Subaru Telescope.

In brief, the image processing builds on \astromatic~software suite\footnote{\url{https://www.astromatic.net}}. In particular, astrometric solution is based on \scamp~\citep{bertin2006,bertin2010} using A1689 exposures but also archival images taken during the same observing run, with a slidding window of several days around the exposures of interest so that at least 3 exposures constitute the same astrometric instrument. The typical internal accuracy is of order $15 {\rm mas}$ using Two Micron All Sky Survey (2MASS) \citep{2mass}. It could definitely improve further with more recent GAIA reference but we deemed it sufficient for our current purpose. \swarp~is used for image coaddition \citep{bertin2002} after the exclusion of non-photometric exposures. The selection is also based on seeing conditions. We also build a model of the Point-Spread Function (PSF) and its spatial variations is performed with \psfex~\citep{bertin2011} for all the exposures in all the bands. A PSF model is also built on the coadded frames. The recovered PSF Full Width at Half Maximum (FWHM) is found to be $0\farcs81$, $0\farcs70$, $0\farcs62$, $0\farcs69$, $0\farcs68$, in the $B$, $V$, $R_\mathrm{C}$, $i^+$, and $z^+$ bands, respectively. Galaxy shapes were measured from the $R_\mathrm{C}$-band data, which provide the highest image quality within the dataset, both in terms of depth and seeing. Source ellipticities were measured using the model-fitting capabilities of \sex, assuming the surface brightness distribution of galaxies can be well approximated by a single Sersic profile. The accuracy of this shape measurement technique was extensively assessed in the GREAT3 challenge as part of the Amalgam team \citep{Mandelbaum2015} and further explored in the context of the preparation of the Euclid mission \citep{EuclidIV}. Residual multiplicative of additive biases are thus kept well below the statistical uncertainties and the weak lensing signal expected from this cluster.

Photometric redshifts (photo-$z$) for individual galaxies were estimated by matching the Subaru $BVR_\mathrm{C}i^+z^+$ photometry to the 30-band photometric redshift (photo-$z$) catalog of the 2~deg$^2$ COSMOS field \citep{laigle16}. We first degrade or {\it perturb} COSMOS2015 photometry when necessary to match the depth of our A1689 data by adding extra noise in the COSMOS2015 catalogue. Then, nearest neighbors in the matched $BVR_\mathrm{C}i^+z^+$ magnitude space are used for each source in order to build a photometric redshift distribution based the 100 nearest neighbors using a fast kd-tree structure  \citep{kdtree2}. The resulting mean, median and dispersion redshift are obtained along with a marginalized mean and standard deviation of the distance ratio $\beta \equiv D_{\rm ls}/D_{\rm s}$. 
A secure selection of background galaxies is key for obtaining accurate cluster mass measurements from weak
lensing. The cluster lensing signal, $\langle g_+(\theta)\rangle$, was measured from a background galaxy sample selected based on a color--color (CC) cut method. This method has been calibrated against evolutionary color tracks of galaxies and photo-$z$ catalogs from deep multiwavelength surveys such as COSMOS \citep[for details, see][]{Medezinski+2010,Medezinski2018src}. In our analysis, CC cuts were performed using $BR_\mathrm{C}z^+$ photometry from Subaru/Suprime-Cam, which offers broad coverage across the optical wavelength range and is well-suited for cluster weak-lensing studies \citep[e.g.,][]{Umetsu2014clash,Umetsu2022}. These CC cuts yield a background sample of 16,428 galaxies, corresponding to a mean surface number density of $n_g\approx 15$~galaxies~arcmin$^{-2}$. For the selected background sample, we find a weighted mean of $\langle \Sigma_\mathrm{\rm crit}^{-1}\rangle^{-1}=3.53\times 10^{15}M_\odot$~Mpc$^{-2}$ and $f_l = \langle\Sigma_\mathrm{\rm crit}^{-2}\rangle/\langle \Sigma_\mathrm{\rm crit}^{-1}\rangle^2 \approx 1.022$.

In Fig.~2, we show the resulting azimuthally averaged $\langle g_+(\theta)\rangle$ profile for the cluster. The profile was measured in 11 logarithmically spaced bins spanning the radial range $\theta\in [2, 20]$~arcmin, corresponding to cluster-centric comoving radii of $(1+z_l)D_l \theta\in [0.3, 3.0]$~$h^{-1}$~Mpc. The error covariance matrix for the binned $\langle g_+(\theta_i)\rangle$ profile is expressed as $C_{ij} = C^\mathrm{shape}_{ij} + C^\mathrm{lss}_{ij}$, where $C^\mathrm{shape}_{ij}$ is the diagonal statistical uncertainty due to the shape noise and $C^\mathrm{lss}_{ij}$ accounts for the cosmic noise due to projected uncorrelated large scale structure \citep{Hoekstra2003}. The components of the $C^\mathrm{lss}$ matrix were evaluated in our fiducial WMAP9 cosmology \citep[see, e.g.,][]{Umetsu2020rev}.

\section{Results}

\begin{table*}
\caption{Combined Analysis Posterior Distributions: Median Values and $1\sigma$ uncertainties (i.e. posterior distribution between the 16th and 84th percentile).}
\label{table:combined}
\centering

\begin{tabular}{ccccccccccc}
\toprule

Model name & $P_{NT}$ type & $c_{200}$ & $r_{200}$ [kpc] & $M_{200}$ [$10^{15}M_{\odot}$] & $\alpha_{\operatorname{NT}}( \frac{r_{200}}{2})$ & $e$ & $c_{\text{norm}}$ & $r_s$ & $\alpha$ \\
\midrule

\multicolumn{9}{c}{NFW X-ray analysis} \\
\midrule

1-a & None & $6.01^{+0.23}_{-0.24}$ & $2141^{+33}_{-29}$ & $1.30^{+0.06}_{-0.05}$ & x & x & x & x & x \\ 

\midrule
\multicolumn{9}{c}{NFW X-ray and SZ analysis} \\
\midrule

1-b  & None & $6.14^{+0.19}_{-0.18}$ & $2188^{+23}_{-22}$ & $1.39^{+0.04}_{-0.04}$ & x & x & x & x & x \\ 
1-c & None & $6.48^{+0.21}_{-0.20}$ & $2039^{+22}_{-23}$ & $1.13^{+0.04}_{-0.04}$ & x & $1.30^{+0.03}_{-0.03}$ & x & x & x \\ 

\midrule
\multicolumn{9}{c}{NFW (1st row) and Einasto (2nd row) weak lensing only analysis} \\
\midrule

2-a& None & $15.40^{+3.03}_{-3.38}$ & $2298^{+95}_{-81}$ & $1.61^{+0.21}_{-0.16}$ & x & x & x & x & x \\ 
2-b & None & $7.52^{+0.67}_{-1.20}$ & $2593^{+144}_{-167}$ & $2.32^{+0.41}_{-0.42}$ & x & x & $4.07^{+0.61}_{-0.81}$ & $344.7^{+88.5}_{-48.7}$ & $0.17^{+0.12}_{-0.06}$ \\

\midrule
\multicolumn{9}{c}{NFW joint X-ray and weak lensing analysis} \\
\midrule

3-a & None & $5.89^{+0.23}_{-0.22}$ & $2165^{+33}_{-30}$ & $1.35^{+0.06}_{-0.06}$ & x & x & x & x & x \\ 
3-b & None & $5.49^{+0.24}_{-0.22}$ & $2154^{+32}_{-29}$ & $1.33^{+0.06}_{-0.05}$ & x & $1.59^{+0.14}_{-0.14}$ & x & x & x \\ 
3-c & Polytropic & $6.60^{+0.72}_{-0.57}$ & $2450^{+81}_{-72}$ & $1.95^{+0.20}_{-0.17}$ & $0.29^{+0.06}_{-0.06}$ & x & x & x & x \\ 
3-d & Angelinelli & $8.21^{+0.97}_{-0.83}$ & $2539^{+77}_{-72}$ & $2.17^{+0.21}_{-0.18}$ & $0.28^{+0.05}_{-0.05}$ & x & x & x & x \\ 
3-e & Polytropic & $5.95^{+0.71}_{-0.47}$ & $2330^{+76}_{-72}$ & $1.68^{+0.17}_{-0.15}$ & $0.19^{+0.07}_{-0.06}$ & $1.35^{+0.18}_{-0.15}$ & x & x & x \\ 
3-f & Angelinelli & $8.40^{+1.53}_{-1.12}$ & $2584^{+186}_{-139}$ & $2.29^{+0.53}_{-0.35}$ & $0.35^{+0.28}_{-0.11}$ & $0.93^{+0.23}_{-0.23}$ & x & x & x \\ 

\midrule
\multicolumn{9}{c}{NFW joint X-ray, SZ, and weak lensing analysis} \\
\midrule

4-a  & None & $6.41^{+0.20}_{-0.22}$ & $2046^{+26}_{-24}$ & $1.14^{+0.05}_{-0.04}$ & x & $1.30^{+0.03}_{-0.03}$ & x & x & x \\
4-b & Polytropic & $6.44^{+0.77}_{-0.56}$ & $2369^{+73}_{-71}$ & $1.76^{+0.17}_{-0.15}$ & $0.34^{+0.05}_{-0.06}$ & $1.30^{+0.03}_{-0.03}$ & x & x & x \\ 
4-c & Angelinelli & $7.07^{+0.67}_{-0.65}$ & $2417^{+74}_{-76}$ & $1.88^{+0.18}_{-0.17}$ & $0.39^{+0.06}_{-0.07}$ & $1.29^{+0.03}_{-0.03}$ & x & x & x \\ 

\midrule
\multicolumn{9}{c}{Einasto joint X-ray, SZ, and weak lensing analysis} \\
\midrule

4-d & Polytropic & $5.83^{+0.51}_{-0.63}$ & $2415^{+106}_{-101}$ & $1.87^{+0.26}_{-0.23}$ & $0.36^{+0.05}_{-0.05}$ & $1.30^{+0.03}_{-0.03}$ & $3.12^{+0.41}_{-0.42}$ & $414.6^{+70.9}_{-49.3}$ & $0.21^{+0.03}_{-0.02}$ \\ 

\bottomrule
\end{tabular}

\end{table*}

\subsection{Gas-only analysis}
\label{sec:gasonly}

\begin{figure*}[htbp]
  \begin{adjustwidth}{}{}
    \centering
    \begin{subfigure}[b]{0.49\textwidth}
      \centering
      \includegraphics[width=\linewidth, trim={0cm 0cm 0cm 0cm},clip]{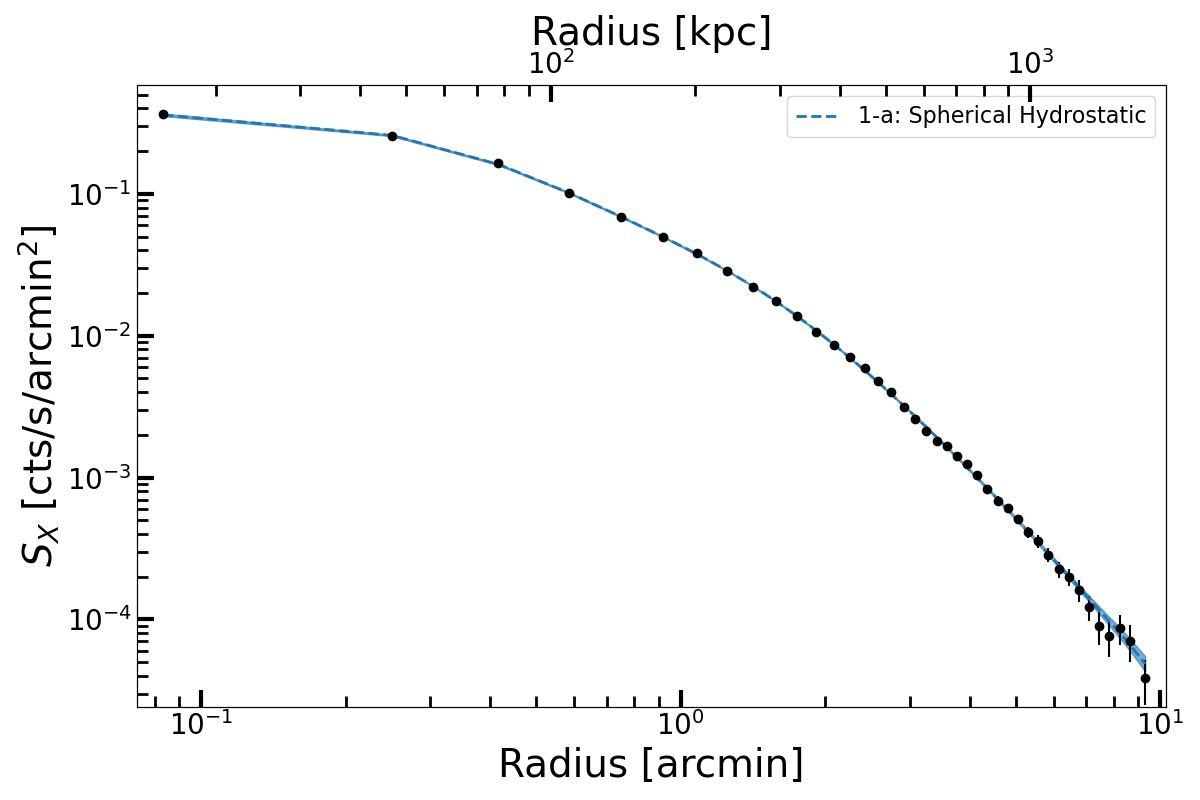}
      \caption{X-ray brightness profile}
      \label{fig:x_only:sx}
    \end{subfigure}
    \hfill
    \begin{subfigure}[b]{0.49\textwidth}
      \centering
      \includegraphics[width=\linewidth, trim={0cm 0cm 0cm 0cm},clip]{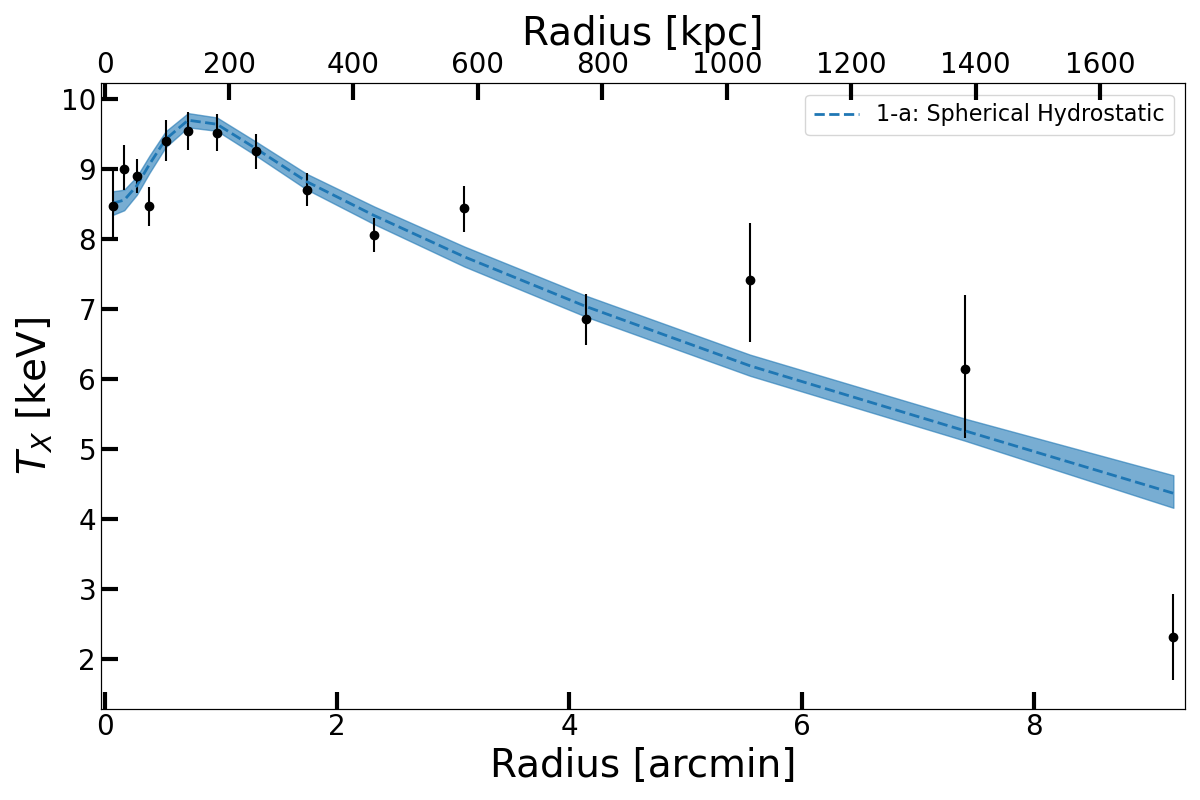}
      \caption{X-ray temperature profile}
      \label{fig:x_only:tx}
    \end{subfigure}
    \caption{A1689 NFW X-ray data fit in the spherical hydrostatic case.}
    
    \label{fig:x_only}
  \end{adjustwidth}
\end{figure*}

Using only the gas observables provided by the X-ray data and assuming both hydrostatic equilibrium and spherical symmetry, we used the mass model method to fit an NFW profile to the X-ray surface brightness and spectroscopic temperature profiles as shown in Fig. \ref{fig:x_only}. The median and the errors of the posterior distributions of the fitted parameters ($c_{200}$ and  $r_{200}$) as well as $M_{200}$ can be found in Table  \ref{table:combined}, where this first analysis can be found under the name "1-a". The recovered concentration $c_{200} = 6.01^{+0.23}_{-0.24}$ and mass $M_{200} = 1.30^{+0.06}_{-0.05} \ 10^{15} M_\odot$ are in good agreement with the earlier X-ray analysis of the \emph{Chandra} data on A1689 \citep{Peng_2009}.

\begin{figure*}[htbp]
  \begin{adjustwidth}{}{}
    \centering
    \begin{subfigure}[b]{0.49\textwidth}
      \centering
      \includegraphics[width=\textwidth, trim={0cm 0cm 0cm 0cm},clip]{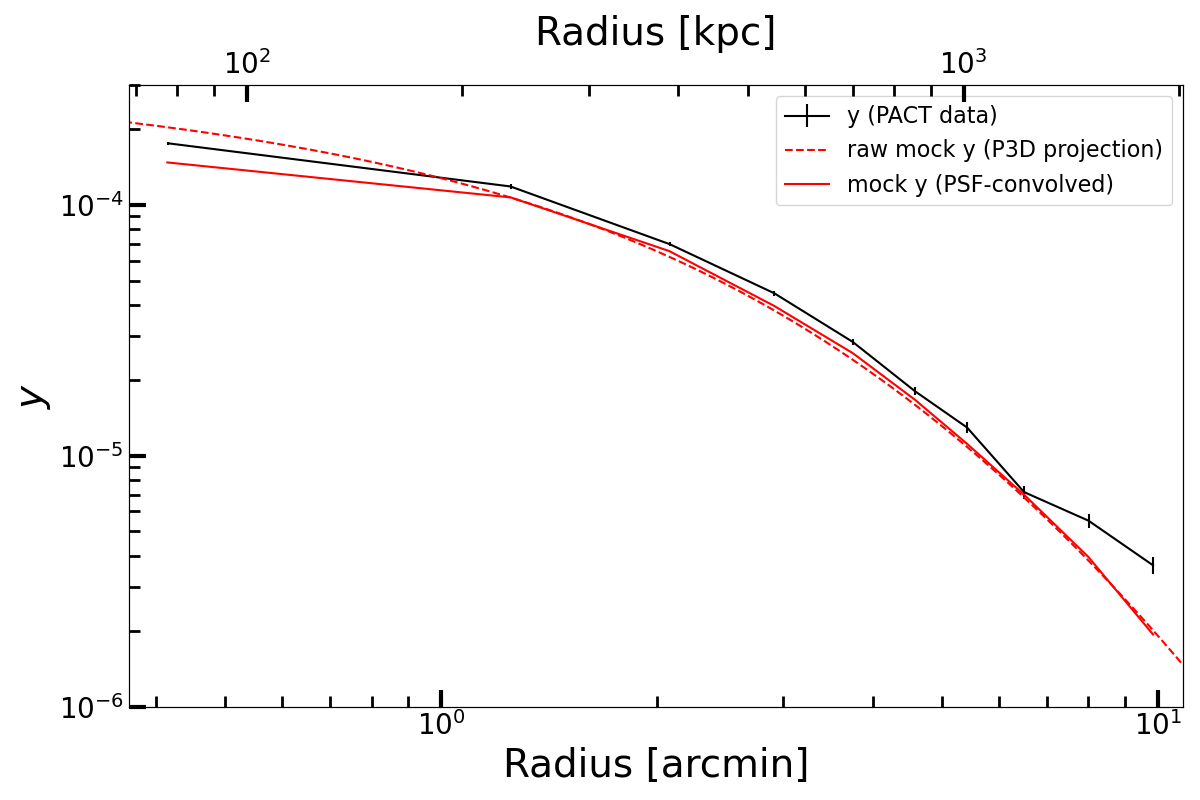}
      \caption{}
      \label{fig:xsz:mock_y}
    \end{subfigure}
    \hfill
    \begin{subfigure}[b]{0.49\textwidth}
      \centering
      \includegraphics[width=\textwidth, trim={0cm 0cm 0cm 0cm},clip]{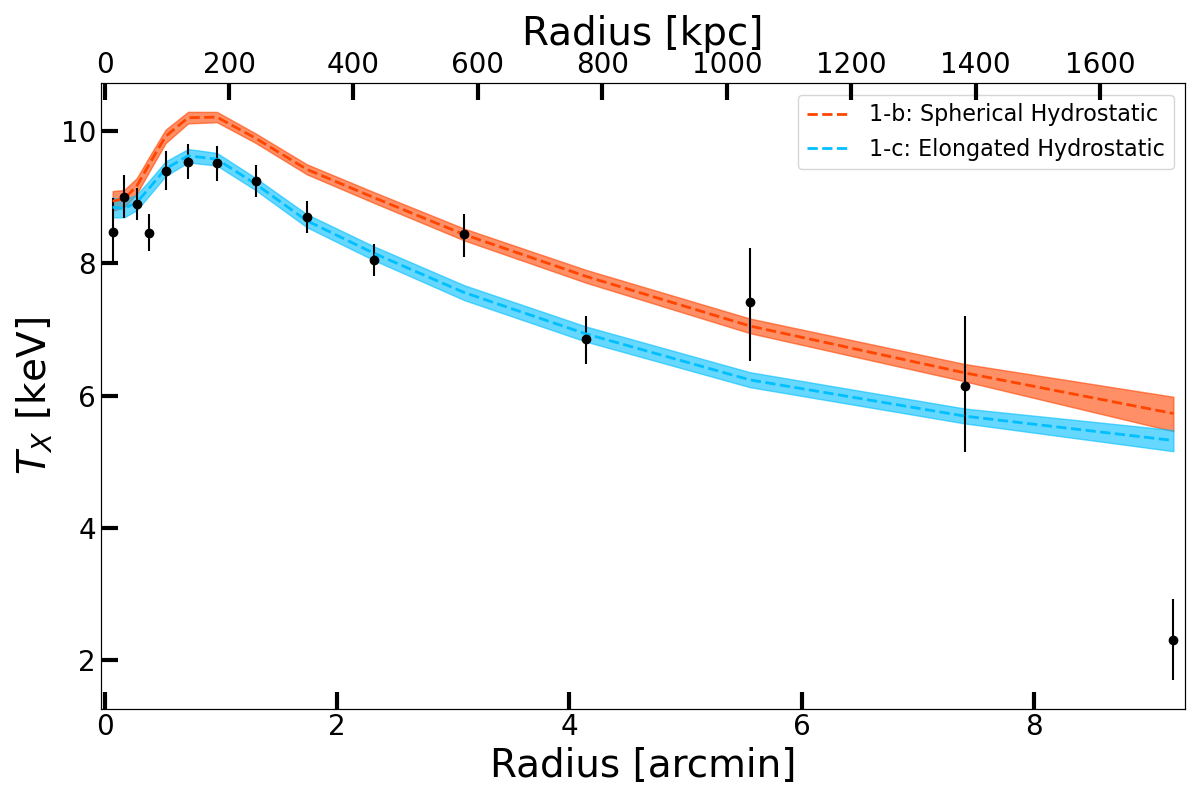}
      \caption{}
      \label{fig:xsz:xsz}
    \end{subfigure}
    \caption{
    a) The gNFW pressure profile fitted to A1689's X-ray data is projected and transformed into a Compton $y$ parameter profile (dashed red) and convolved with the $Planck$ + ACT PSF to produce a realist mock observation (red). This profile is overplotted to the real $y$ profile (black) extracted from $Planck$ + ACT data..\\
    b) NFW X-ray/SZ joint analysis of A1689, X-ray temperature profile posterior distributions of 2 models: spherical and elongated (respectively in orange and light blue, model names 1-b and 1-c in Table \ref{table:combined}). The results of the analysis as a posterior model distribution on the SZ $y$ parameter and X-ray brightness profiles for both models can be found in the appendix.}
    
    \label{fig:xsz}
  \end{adjustwidth}
\end{figure*} 

To test for consistency between the X-ray and SZ profiles, we attempted to compare the pressure profiles inferred from the two methods under the assumption of spherical symmetry. To this aim, we used the parametric forward method (see Sect. \ref{sec:mass_modeling}) to fit a gNFW pressure profile (Eq. \ref{eq:gnfw_pressure}) to the X-ray data. A mock Compton-$y$ parameter profile was inferred by projecting the fitted pressure profile using Eq. \ref{eq:sz}. Convolving this raw mock profile with the PSF model used for the extraction of the $y$ profile of the \emph{Planck} + ACT SZ data allows a fair comparison between our X-ray inferred mock $y$ profile and the actual \emph{Planck} + ACT $y$ profile (see Fig. \ref{fig:xsz:mock_y}). We can see that the mock $y$ profile inferred from the X-ray significantly under-predicts the observed SZ $y$ profile. This suggests that the cluster is elongated along the LOS. Indeed, the SZ signal and the X-ray brightness increase linearly with elongation (Eq. \ref{eq:elongprior}), such that the SZ-inferred pressure scale linearly with $e$, whereas the X-ray-inferred pressure scales with the square root of the elongation. It is thus expected that an object elongated along the LOS will exhibit a lower X-ray pressure than its SZ counterpart. We used {\tt PyMC} to fit for the mean ratio of the two profiles, $y_{\operatorname{Planck + ACT}} = y_{\operatorname{X, mock}} \times e^{1/2}$, accounting for the uncertainties in the SZ data. We obtain $e = 1.28 \pm 0.02$, which significantly exceeds the value of $1.0$ expected in spherical symmetry.

Adding the Compton $y$ parameter profile in addition to the X-ray brightness and temperature profiles, we used the mass model method to perform a joint X-ray and SZ NFW analysis. The two fitted models, spherical hydrostatic (model name 1-b) and elongated hydrostatic (model name 1-c), are shown in Fig. \ref{fig:xsz:xsz} as a fit to the X-ray temperature profile. The median and confidence interval of the best-fitting posterior distributions for both models (with $e$ the elongation parameter in addition to $c_{200}$ and  $r_{200}$ in the elongated case) as well as the inferred $M_{200}$ values can be found in Table \ref{table:combined}. 

As anticipated from the comparison between the gNFW fit of the X-ray data and the $y$ parameter profile (Fig. \ref{fig:xsz:mock_y}), while both models successfully reproduce the X-ray brightness, the spherical model under-predicts the observed $y$ profile (Fig. \ref{ap:xsz}). It also systematically overpredicts the X-ray temperature profile (see Fig. \ref{fig:xsz:xsz}), which implies that deviations from spherical symmetry must be taken into account to accurately model the system. Conversely, the elongated model shows elongation along the LOS at high significance ($e = 1.30^{+0.03}_{-0.03}$), which allows the model to adequately fit both the X-ray and the SZ data (Fig. \ref{fig:xsz:xsz}). Our result is in agreement with \cite{chexmate_2023}, where a 3D triaxial modeling of the cluster was fitted to 2D maps of the same data set of gas observables (X-ray and SZ), using parameterized electronic density and gas pressure profiles and without hydrostatic assumption, resulting in an equivalent elongation parameter of $e = 1.37^{+0.04}_{-0.04}$ (see derivations in Appendix \ref{sec:kim2024}).

\subsection{Weak lensing only analysis}
\label{sec:wlonly}

\begin{figure}[htbp]

      \centering
      \includegraphics[width=\linewidth, trim={0cm 0cm 0cm 0cm},clip]{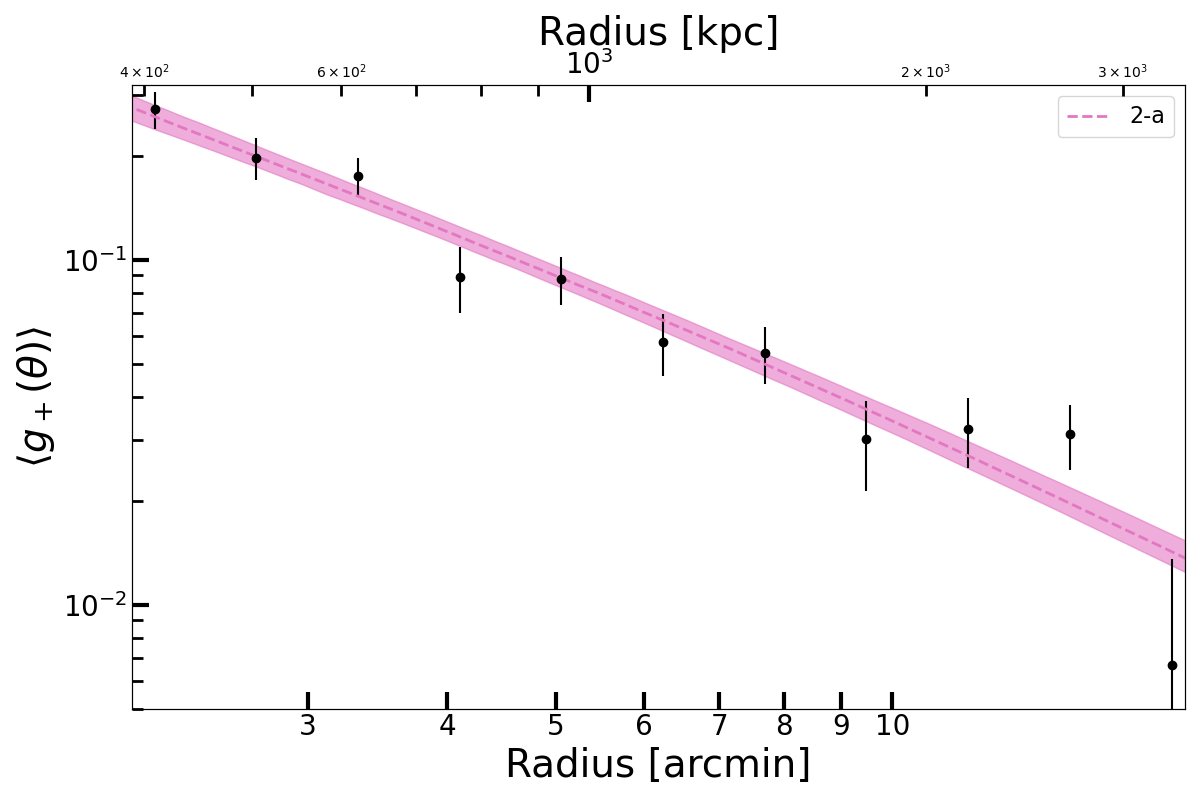}

    \caption{NFW fit of the mean tangential shear profile of A1689 using the weak lensing data only and under spherical symmetry (model 2-a in Table \ref{table:combined}). The black data points show the Subaru/HSC reduced tangential shear profile, whereas the pink curve and shaded area represent the best-fit model and 68\% error envelope.}
    
    \label{fig:wlonly_unifVSgauss}
\end{figure}

When applied to A1689, fitting our numerical model of the weak lensing signal leads to discrepant results with the gas analysis. Assuming spherical symmetry, we fitted the mean tangential shear profile of A1689 by projecting an NFW density profile and computing the reduced shear profile. The NFW model parameters were optimized in {\tt PyMC} assuming Gaussian errors and taking the large-scale structure covariance matrix into account (see Sect. \ref{sec:wl}). 
The posterior envelope of the mean tangential shear profile is shown in Fig. \ref{fig:wlonly_unifVSgauss}, where the dashed lines show the median posterior model curves. The fitted parameter values can be found in Table \ref{table:combined}.

The very steep tangential shear profile of A1689 leads the analysis to very high concentration ($c_{200} = 15.40^{+3.03}_{-3.38}$). The inferred mass ($M_{200} = 1.61^{+0.21}_{-0.16} \ 10^{15}M_\odot$) exceeds the mass inferred from the gas-only analysis (see model 2-a in Table \ref{table:combined}) by 20\%. These results converge towards the already well-known discrepancy between gas observables and weak lensing for this specific cluster \citep{Umetsu_2008, Peng_2009}. Our results are indeed in agreement with the previous weak lensing analysis of A1689 by \cite{Umetsu_2008} ($c_{200} = 10.7^{+4.5}_{-2.7}$, $M_{200}=1.76\pm0.20$).

\subsection{Joint X-ray/weak lensing analysis}
\label{sec:XWL}

As discussed in Sect. \ref{sec:model_NTpressure}, jointly fitting gas and WL data requires the inclusion of a non-thermal pressure term to bridge a potential gap between WL and gas observables. The addition of non-thermal pressure allows to increase the system mass without impacting the fit to the gas observables (Eq. \ref{eq:pnt}). In this section, NFW profiles are jointly fitted to A1689 X-ray and WL data using the mass model method. We compare 6 models that exhaustively summarize all combinations of non-thermal pressure parameterizations and elongation allowed in our framework:

\begin{itemize}
    \item 3-a: spherical hydrostatic,
    \item 3-b: elongated hydrostatic,
    \item 3-c: spherical and polytropic non-thermal pressure,
    \item 3-d: spherical and Angelinelli non-thermal pressure,
    \item 3-e: elongated and polytropic non-thermal pressure,
    \item 3-f: elongated and Angelinelli non-thermal pressure.
\end{itemize}

    The medians and errors of the NFW fitted parameters posteriors ($c_{200}$ and $r_{200}$) and the corresponding enclosed mass $M_{200}$ for the X-ray and weak lensing joint analysis are shown in Table \ref{table:combined} for all 6 models, as well as the elongation $e$ and the non-thermal to thermal pressure ratio $\alpha_{\operatorname{NT}}(\frac{r_{200}}{2})$ (see Eq. \ref{eq:alpha}) when it applies. The ratio $\alpha_{\operatorname{NT}}(\frac{r_{200}}{2})$ was picked as a representative quantity of the non-thermal pressure to offer a fair comparison regardless of the parameterization. The radial distance $\frac{r_{200}}{2}$ being covered by the entirety of the dataset indeed offers a good level of constraints while being at sufficiently large radii to expect an important contribution of the non-thermal pressure to the total pressure \citep{Angelinelli_2020,Nelson_2014}. 

\begin{figure*}[htbp]
  \begin{adjustwidth}{}{}
    \centering
    \begin{subfigure}[b]{0.49\textwidth}
      \centering
      \includegraphics[width=\linewidth, trim={0cm 0cm 0cm 0cm},clip]{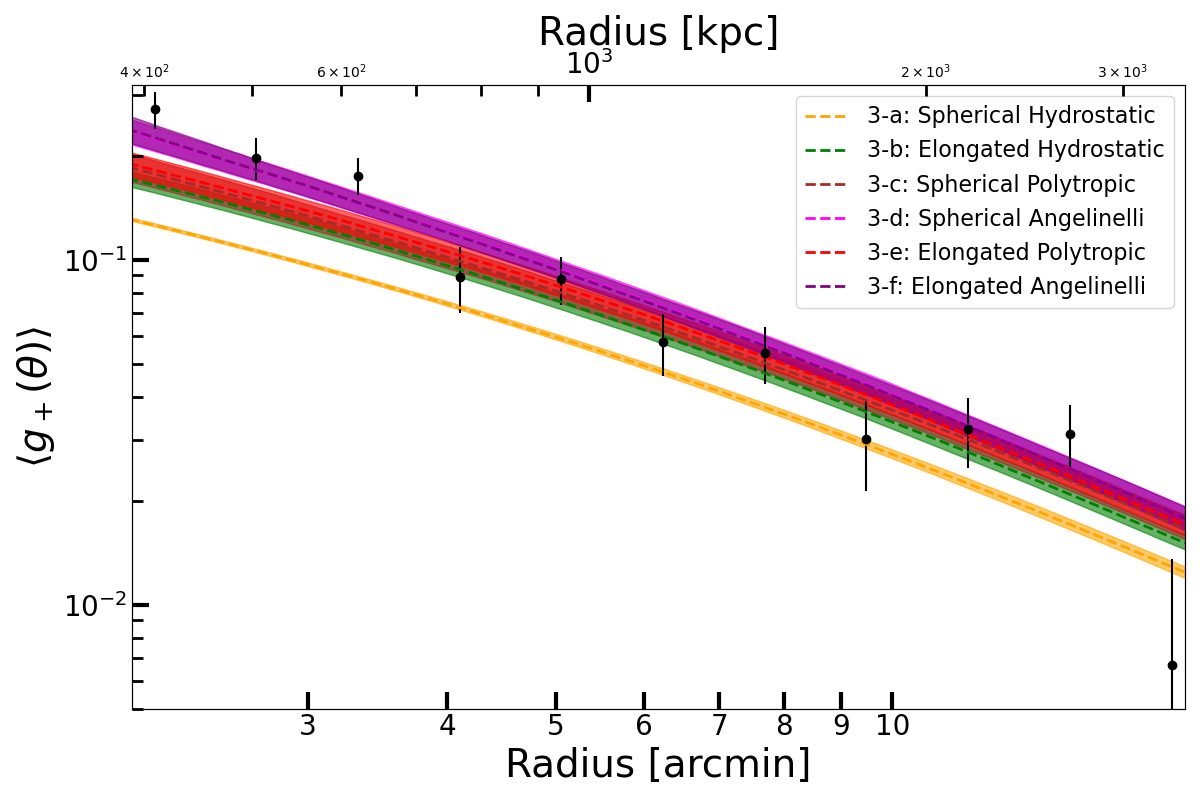}
      \caption{}
      \label{fig:gpluscompare:xwl}
    \end{subfigure}
    \hfill
    \begin{subfigure}[b]{0.49\textwidth}
      \centering
      \includegraphics[width=\linewidth, trim={0cm 0cm 0cm 0cm},clip]{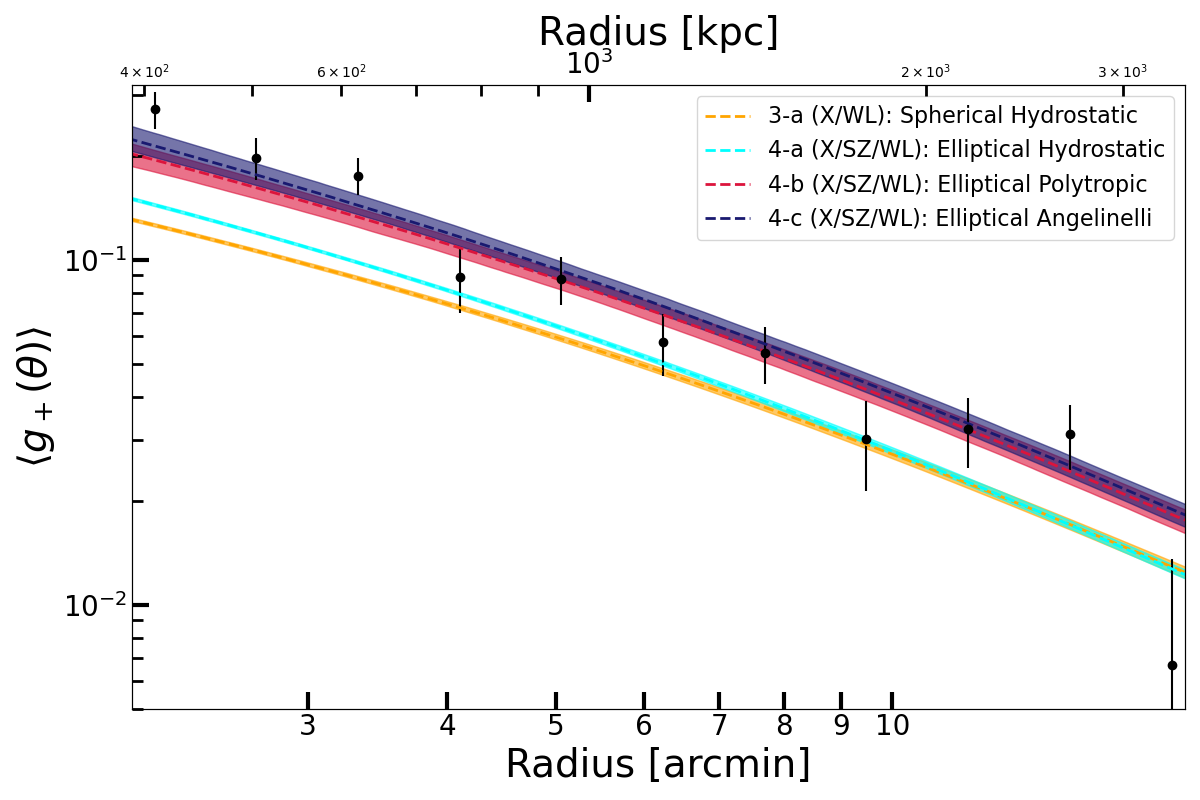}
      \caption{}
      \label{fig:gpluscompare:xszwl}
    \end{subfigure}
    \caption{A1689 NFW joint analysis, fit to the tangential shear profile (black points). The corresponding fits to the gas observable can be found in the appendix. \emph{(a)} X-ray/WL analysis results. The X-ray 1D profiles (surface brightness and temperatures) as well as the tangential shear profile are fitted jointly for every $P_{NT}$ (Angelinelli or polytropic parameterization) and elongation combination (models 3-a through 3-f; see Sect. \ref{sec:XWL} and Table \ref{table:combined}). \emph{(b)} X-ray/SZ/WL analysis results (see Sect. \ref{sec:full}). All gas observable profiles ($S_X$, $T_X$, Compton $y$) are fitted jointly with the tangential shear profile. The three X-ray/SZ/WL results shown here include the elongation parameter (i.e. elongated models) and are either not integrating any non-thermal pressure (4-a hydrostatic; cyan) or one of our two models (4-b polytropic, red; 4-c Angelinelli, purple). We also display the simple X-ray/WL joint analysis (3-a; orange) for a visual comparison with the spherical case.}
    
    \label{fig:gpluscompare}
  \end{adjustwidth}
\end{figure*}

In Fig. \ref{fig:gpluscompare:xwl} we show all 6 posterior model envelopes for the joint X-ray and WL NFW analysis of A1689 as a fit to the mean tangential shear. The dashed lines show the median posterior model curves. The results of the analysis, displayed as the envelopes of the X-ray gas observables, are shown in Fig. \ref{ap:xwl}. The X-ray brightness $S_X$ and temperature $T_X$ profiles are always correctly reproduced by all models. The high statistical power of the X-ray data as compared to the weak lensing signal allows X-rays to drive the fit in the case where the model cannot provide a good fit to both, as shown by the simple spherical hydrostatic model in orange (model 3-a). Indeed, while failing at reproducing the weak lensing signal, it still provides a good prediction of the X-ray data (see Fig. \ref{ap:xwl}). By systematically underpredicting the shear, and similarly to the conclusion made after our X-ray and SZ joint fit in the previous section, model 3-a shows the limitation of the spherical hydrostatic NFW. On the other hand, either the non-thermal pressure - independently of the chosen model - or the elongation, as well as their combination, reproduces the shear signal efficiently, while keeping a similarly good fit to the X-ray data. 

\begin{figure*}[htbp]
    \centering
    \begin{subfigure}[b]{0.48\textwidth} 
        \centering
        \includegraphics[width=\linewidth]{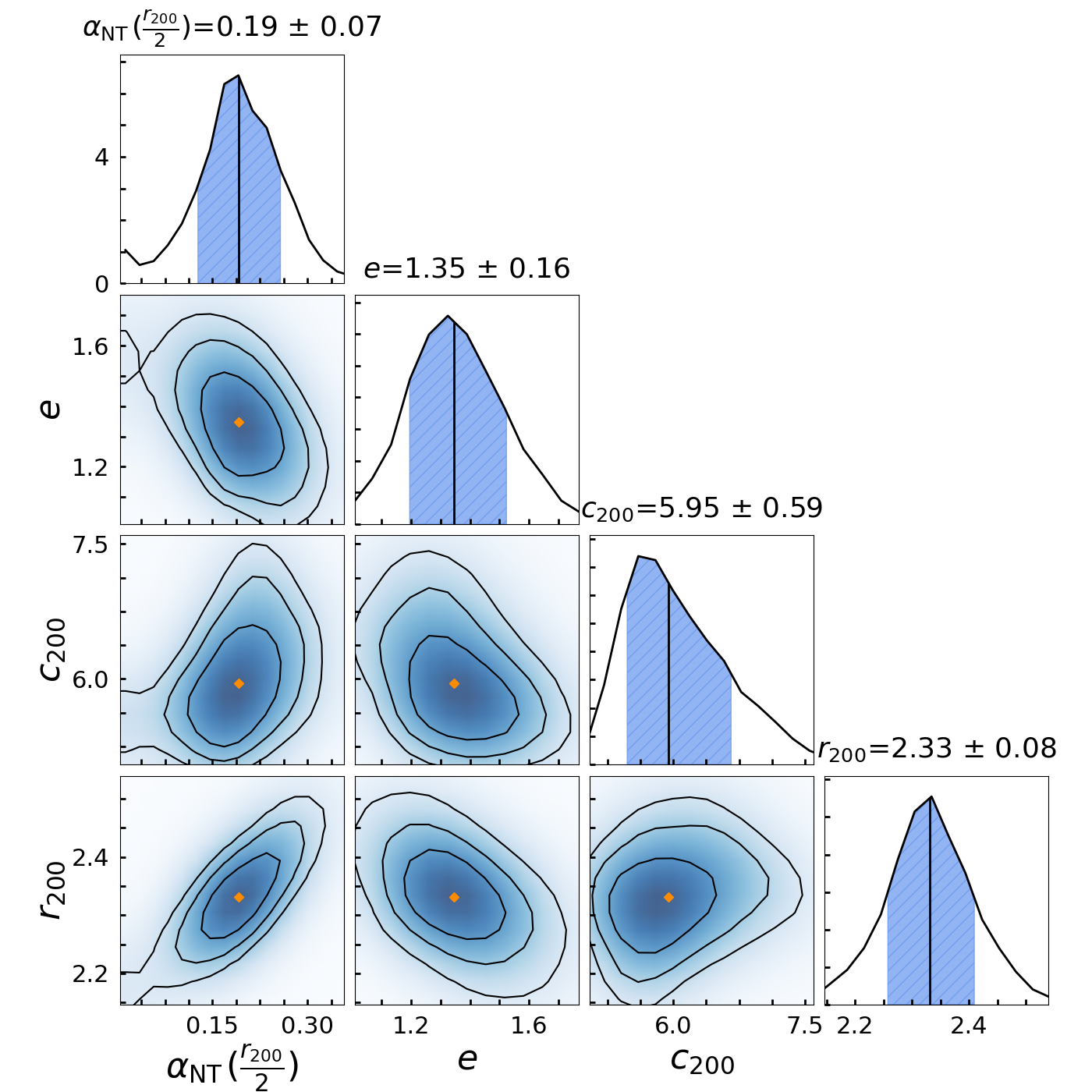}
        \caption{}
        \label{fig:pntVSelong_poly:xwl}
    \end{subfigure}
    \hfill
    \begin{subfigure}[b]{0.48\textwidth} 
        \centering
        \includegraphics[width=\linewidth]{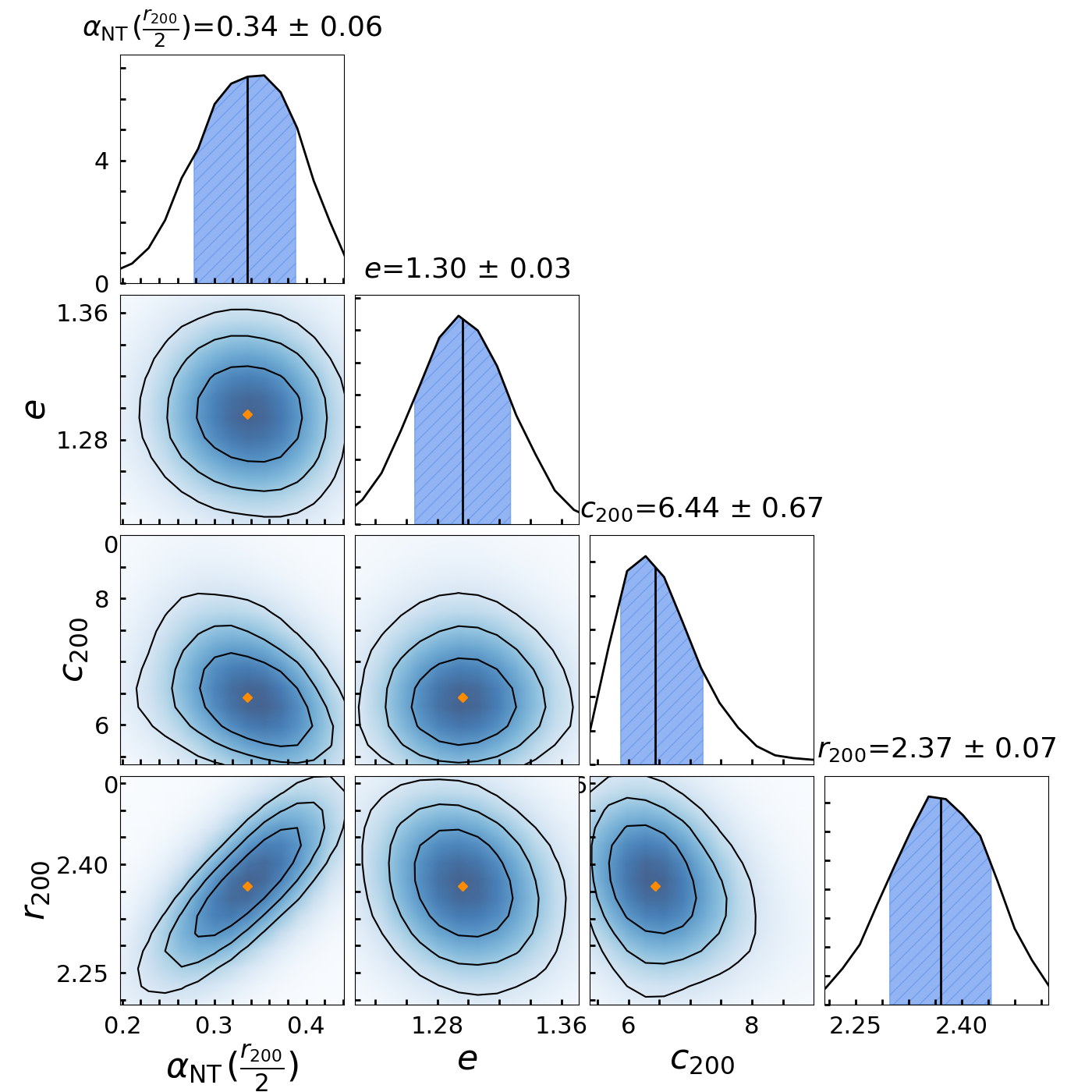}
        \caption{}
        \label{fig:pntVSelong_poly:xszwl}
    \end{subfigure}
    
    \caption{Corner plots of the main model parameters for the NFW joint analysis of A1689 for the polytropic non-thermal pressure model with elongation (Eq. \ref{eq:pnt_ettori}). The displayed model parameters are the non-thermal pressure ratio at a radius of $0.5r_{200}$ ($\alpha_{\operatorname{NT}}(0.5r_{200})$), the LOS elongation $e$, the concentration $c_{200}$, and the overdensity radius $r_{200}$ (in Mpc). Panel (a) shows the corner plot for the X-ray/WL analysis (model 3-e), whereas panel (b) shows the posteriors for the full X-ray/SZ/WL joint analysis (model 4-b).}
    \label{fig:pntVSelong_poly}
\end{figure*}

In this joint X-ray/WL analysis, the non-thermal pressure and elongation play a very similar role. By shifting the reconstructed shear up, these two parameters allow the model to fit both the weak lensing and the X-ray data. As a direct consequence, elongation and non-thermal pressure are here strongly anti-correlated. This is illustrated for the polytropic parameterization of the non-thermal pressure in Fig. \ref{fig:pntVSelong_poly:xwl}, showing the model 3-e corner plot: non-thermal pressure to total pressure ratio $\alpha_{\operatorname{NT}}$ versus the elongation parameter. The same results are observed with the Angelinelli parameterization (model 3-f) in Fig. \ref{ap:pntVSelong_ang:xwl}. 

We note that all considered models imply a concentration in the range 6-8, which is much lower than the concentration retrieved in the WL-only case (see Sect. \ref{sec:wlonly}). The addition of the X-ray data brings strong constraints on the shape of the underlying mass profile, which drives the concentration towards a lower and probably more realistic value for such a massive halo.

\subsection{Full NFW X-ray/SZ/weak lensing joint analysis: non-thermal pressure parameterization comparison}
\label{sec:full}

As shown by comparing the projected X-ray inferred pressure profile to the SZ data and by doing the joint X-ray/SZ mass model analysis, the pressure profiles inferred from the X-ray and the SZ data have a different LOS dependency (see Sect. \ref{sec:gasonly}). As a direct consequence, the elongation parameter $e$ propagates differently to these two observables, and fitting them jointly puts a tight constraint on the elongation parameter $e$. Applying our framework to the entire data set (X-ray, SZ, and weak lensing), we therefore only consider the elongated scenario and compare three different models assuming the NFW mass model:

\begin{itemize}
    \item 4-a: elongated hydrostatic,
    \item 4-b: elongated and polytropic non-thermal pressure,
    \item 4-c: elongated and Angelinelli non-thermal pressure.
\end{itemize}

The results of the optimization for the three cases above are provided in Table \ref{table:combined}. The fits to the tangential shear modeling of models 4-a, 4-b and 4-c are displayed in Fig. \ref{fig:gpluscompare:xszwl}, whereas the corresponding fits to the gas observables can be found in Fig. \ref{ap:xszwl}. While the X-ray and SZ gas observables are well-fitted by all models without significant differences, the absence of non-thermal pressure (model 4-a) results in a systematic under-prediction of the weak lensing signal. On the other hand, the two models featuring non-thermal pressure support  (models 4-b and 4-c) successfully reproduce all the profiles of the data set.

The corner plot shown in in Fig. \ref{fig:pntVSelong_poly:xszwl} displays the posterior distribution of the elongation parameter, the non-thermal pressure ratio $\alpha_{\operatorname{NT}}$, the NFW concentration and $r_{200}$ for the polytropic parameterization (model 4-b). Jointly fitting the SZ data together with the X-ray and WL data allowed to successfully lift the degeneracy between non-thermal pressure and elongation, as compared with the previous X-ray/WL joint analysis (model 3-e; see Fig. \ref{fig:pntVSelong_poly:xwl}). We obtain the same results for the Angelinelli parameterization of the non-thermal pressure, as shown in Appendix Fig. \ref{ap:pntVSelong_ang:xszwl}. Compared to the X-ray/WL analysis (Sect. \ref{sec:XWL}), adding the SZ data to the joint analysis results in a much tighter constraint on the elongation parameter. 
Consequently, the elongation parameter remains very stable despite changes in the non-thermal pressure parameterization, as its median posterior values, $e = 1.30\pm{0.03}$ remain unchanged for every model (4-a, 4-b and 4-c), and consistent with our gas-only analysis (Sect. \ref{sec:gasonly}) and our previous results considering X-ray and SZ data in a full triaxial model \citep{chexmate_2023}.

As a consequence of the tight constraint set by the addition of the SZ data on the LOS elongation, deviations from spherical symmetry alone cannot reconcile the gas observables with the measured weak lensing signal (see model 4-a in Fig. \ref{fig:gpluscompare:xszwl}). Non-thermal pressure is now the only parameter of our framework that remains free to boost the weak lensing signal by increasing the mass of the system without impacting on the gas observables. Therefore, in the case where high-quality constraints on X-ray, WL, and SZ signals can be obtained, the degeneracy between LOS elongation and non-thermal pressure can be broken and the two parameters can be determined independently (see Fig. \ref{fig:pntVSelong_poly:xszwl}). The non-thermal pressure at $\frac{r_{200}}{2}$ respectively converges to $\sim30\text{-}40\%$ for the polytropic and Angelinelli parameterizations, while keeping a high level of agreement with each other. 

The best-fitting mass model retrieved from the full analysis is fully consistent between the two non-thermal pressure parameterizations. We retrieve a concentration of $\sim7$ and a total spherical mass $M_{200}=1.76\times10^{15}M_\odot$ (with the polytropic parameterization, model 4-b) and $M_{200}=1.88\times10^{15}M_\odot$ (with the Angelinelli parameterization, model 4-c) correcting for LOS elongation. Given the scatter of $\sim40\%$ in the mass-concentration relation \citep[e.g.][]{Diemer_2015,Klypin_2016}, the fitted concentration better agrees with the expected values in $\Lambda$CDM than the value of $\sim15$ retrieved from the WL-only analysis (see Sect. \ref{sec:wlonly}).

\section{Discussion}

\subsection{Pressure profiles}

Unlike the case of the X-ray/WL analysis (Sect. \ref{sec:XWL}), our multi-probe approach, which combines X-ray, SZ, and WL data (Sect. \ref{sec:full}) allows us to break the degeneracy between non-thermal pressure and LOS elongation when comparing mass reconstructions from the ICM and weak gravitational lensing. Our constraints on the two non-thermal pressure parameterizations are of comparable quality. This outcome highlights the robustness of our modeling and supports the polytropic model by bringing its constraints in line with those of the Angelinelli parameterization.

\begin{figure*}[htbp]
  \begin{adjustwidth}{}{}
    \centering
    \begin{subfigure}[b]{0.49\textwidth}
      \centering
      \includegraphics[width=\linewidth, trim={2cm 0cm 4cm 0cm},clip]{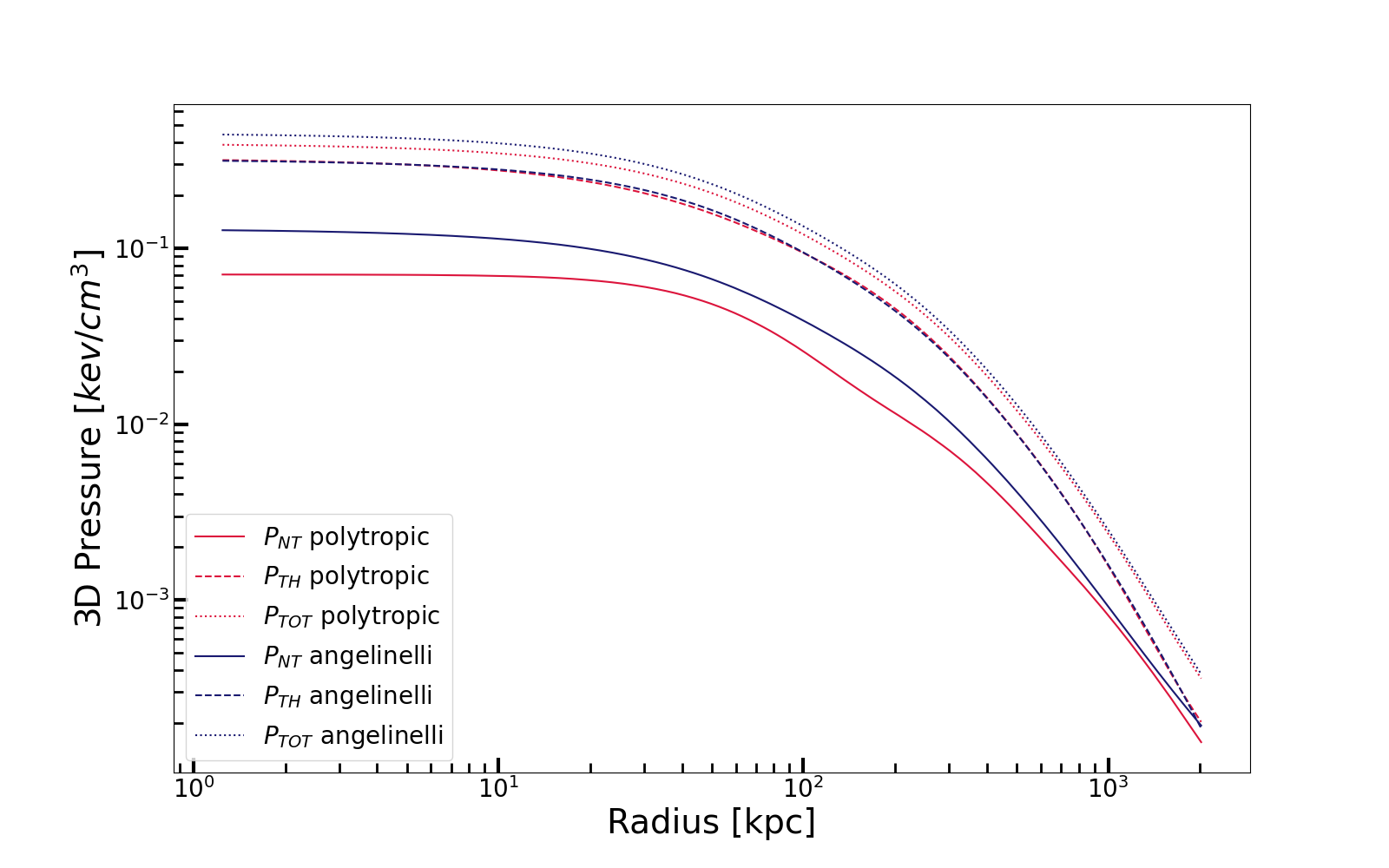}
      \caption{}
      \label{fig:pnt_profiles:prof}
    \end{subfigure}
    \hfill
    \begin{subfigure}[b]{0.49\textwidth}
      \centering
      \includegraphics[width=\linewidth, trim={2cm 0cm 4cm 0cm},clip]{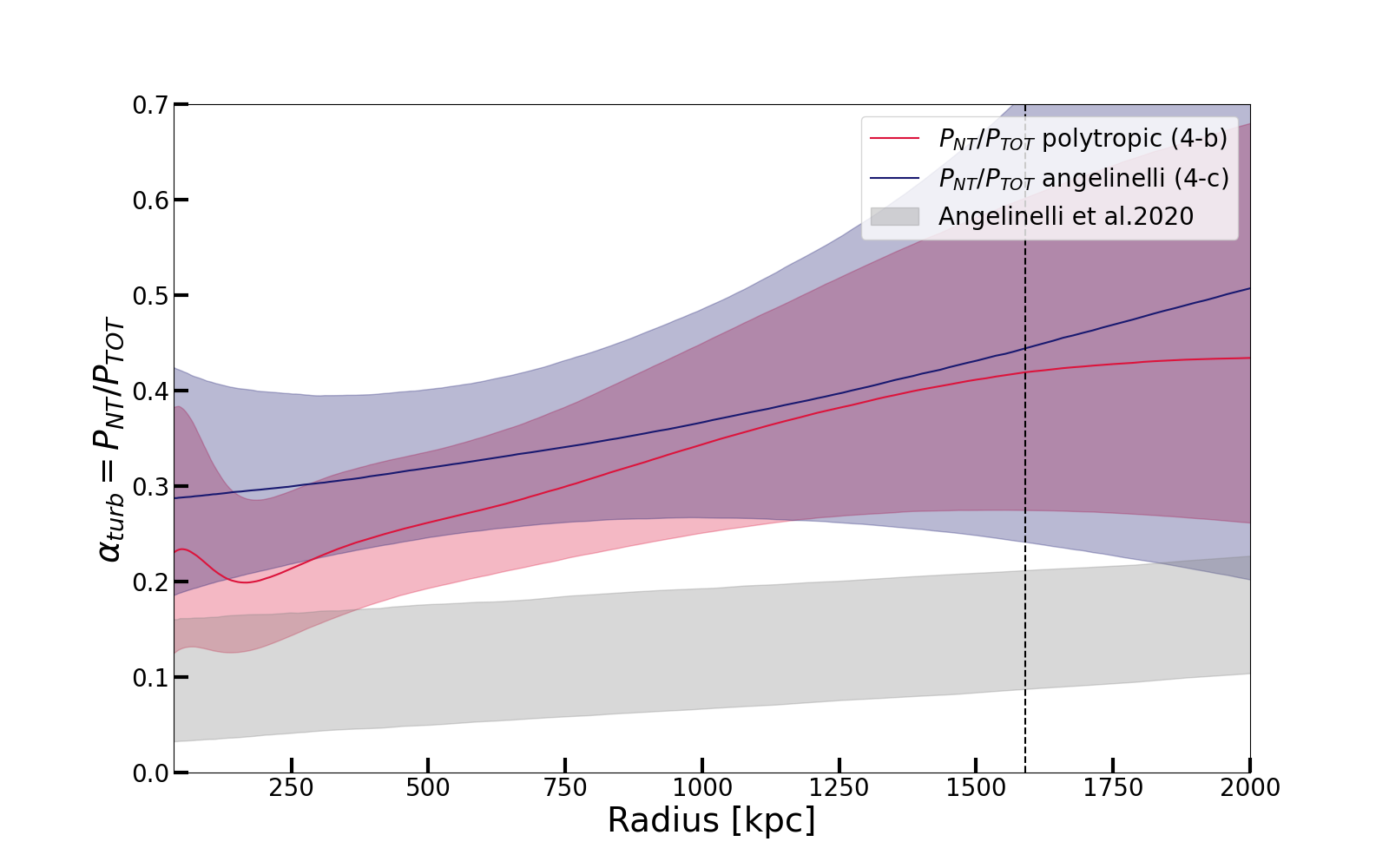}
      \caption{}
      \label{fig:pnt_profiles:ratio}
    \end{subfigure}
    \caption{a) Pressure profiles inferred from the A1689 NFW full joint analysis (X-ray/WL/SZ) of both the polytropic and Angelinelli parameterizations.\\b) $P_{NT}/P_{TOT}$ ratio for both parameterizations. The grey shaded region denotes the 1$\sigma$ envelope of the non-thermal pressure profiles derived from 20 simulated clusters in \citet{Angelinelli_2020}, which were used to inform our prior distributions. The dashed black line marks the radius $r_{500}$.\\The radii considered here are the 3D averaged radii.}
    
    \label{fig:pnt_profiles}
  \end{adjustwidth}
\end{figure*}

The recovered pressure profiles are presented in Fig. \ref{fig:pnt_profiles}. The left panel illustrates the comparison between the thermal and non-thermal pressure profiles and the total pressure derived from the mass model (Eq. \ref{eq:pnt}). At $r_{500}$, the Angelinelli and Polytropic parameterizations indicate non-thermal pressure fractions of approximately and $\alpha_{\operatorname{NT}}\approx 40\%$. These values position A1689 among clusters with significant non-thermal pressure support, notably exceeding the $20\%$ levels predicted by hydrodynamical simulations \citep[e.g.,][]{Rasia_2006,Lau_2009,Nelson_2014,Biffi_2016,Angelinelli_2020,Gianfagna_2023} and the even lower level reported on a set of 12 clusters of the X-COP sample \citet{Dupourque_2023} and 64 CHEX-MATE clusters \citet{Dupourque_2024}, on which the average non-thermal pressure support at $r_{500}$ was found to be not higher than 10$\%$.  

Nonetheless, the envelopes of $\alpha_{\operatorname{NT}}$, shown in the right panel of Fig. \ref{fig:pnt_profiles}, remain broadly consistent with simulation predictions. This aligns with findings by \citet{Morandi_2011}, who reported a $20\%$ non-thermal pressure fraction in A1689 (under the assumption that $\alpha_{\operatorname{NT}}$ is constant with radius). Additionally, our results agree closely with \citet{Umetsu_2015}, who similarly reported a non-thermal pressure contribution reaching $\sim 40\%$ at 0.9Mpc. 

Both non-thermal pressure fraction profiles show a slight increase with radius, this behavior is expected from numerical simulations \citep{Rasia_2006, Lau_2009, Pearce_2019, Angelinelli_2020}, although with significant differences from one system to another. While our results are in qualitative agreement with this prediction, application of our technique to a larger sample of clusters is needed to study the shape of the mean non-thermal pressure profile, as done for the CLASH sample \citep{Sayers_2021}.



\subsection{Einasto density profile}
\label{sec:einasto}

Using our multi-observable framework, we compare the results of our mass modeling between the NFW parameterization used so far to the results obtained with the Einasto density profile. We applied our reconstruction scheme to the full dataset (X-ray/SZ/WL) with the polytropic non-thermal pressure parameterization and allowing for elongation (model 4-d in Table \ref{table:combined}). The posterior distributions of the fit with the Einasto mass model are shown in Appendix Fig. \ref{ap:einasto}. Comparing the resulting mass model to its equivalent NFW  X-ray/SZ/WL multi-probe analysis with polytropic $P_{NT}$ and ellipticity (model 4-b; see the fit to the shear in Fig. \ref{fig:gpluscompare:xszwl} and to the gas observables in Fig. \ref{ap:xszwl}), we can see that the Einasto and NFW density profiles provide an equally good fit to the weak lensing data and the gas observables, without significant differences. 

The median NFW parameters ($c_{200}$ and $r_{200}$) align well with their counterparts derived from the Einasto full joint analysis, as presented in Table \ref{table:combined}, where the Einasto concentration is inferred through the relation $c_{200} = r_{200}/r_{s}$. This agreement is anticipated, considering our Einasto analysis yields a value of the shape parameter $\alpha = 0.21\pm{ 0.03}$, when equivalence with the NFW parameterization is expected for $\alpha \sim 0.2$ \citep{Dutton_2014}. Moreover, \cite{Eckert_2022} measured a mean value of $\alpha=0.19\pm0.03$ with X-ray and SZ data on 12 massive clusters of the XMM-Newton Cluster Outskirts Project. Our best fit for the Einasto index therefore agrees with the expectations from the CDM paradigm \citep[e.g.][]{Navarro_2004,Ludlow_2016,Brown_2020}.

We compared our Einasto multi-probe analysis to a fit to the weak lensing data only using the same model. The WL-only fit to the shear profile is shown in Appendix Fig. \ref{ap:einasto:gt}. In Fig. \ref{fig:alpha_prior_posterior} we show the posterior distributions on the Einasto index $\alpha$ for the multi-probe analysis and the WL data alone. For comparison, we also show the prior distribution on the Einasto shape parameter. The weak lensing only analysis does not add much constraint on the shape parameter, as the posterior distribution is nearly identical to the prior. Conversely, the joint analysis allows us to constrain the Einasto shape parameter $\alpha$, as shown by the narrower posterior distribution and errors (Table \ref{table:combined}), where we can see that the posterior distribution on $\alpha$ significantly pulls away from the prior. The addition of the X-ray and SZ data thus allows us to add further constraints on the shape of the DM density profile and extend beyond the NFW paradigm, which is beyond the reach of WL data even when high-quality data are available. Application of our technique to larger cluster samples will allow us to test CDM predictions on the internal structure of DM halos and provide important constraints on alternative DM scenarios such as self-interacting DM \citep{Robertson_2021,Eckert_2022}.

\begin{figure}[htbp]
\centering
\includegraphics[width=\hsize]{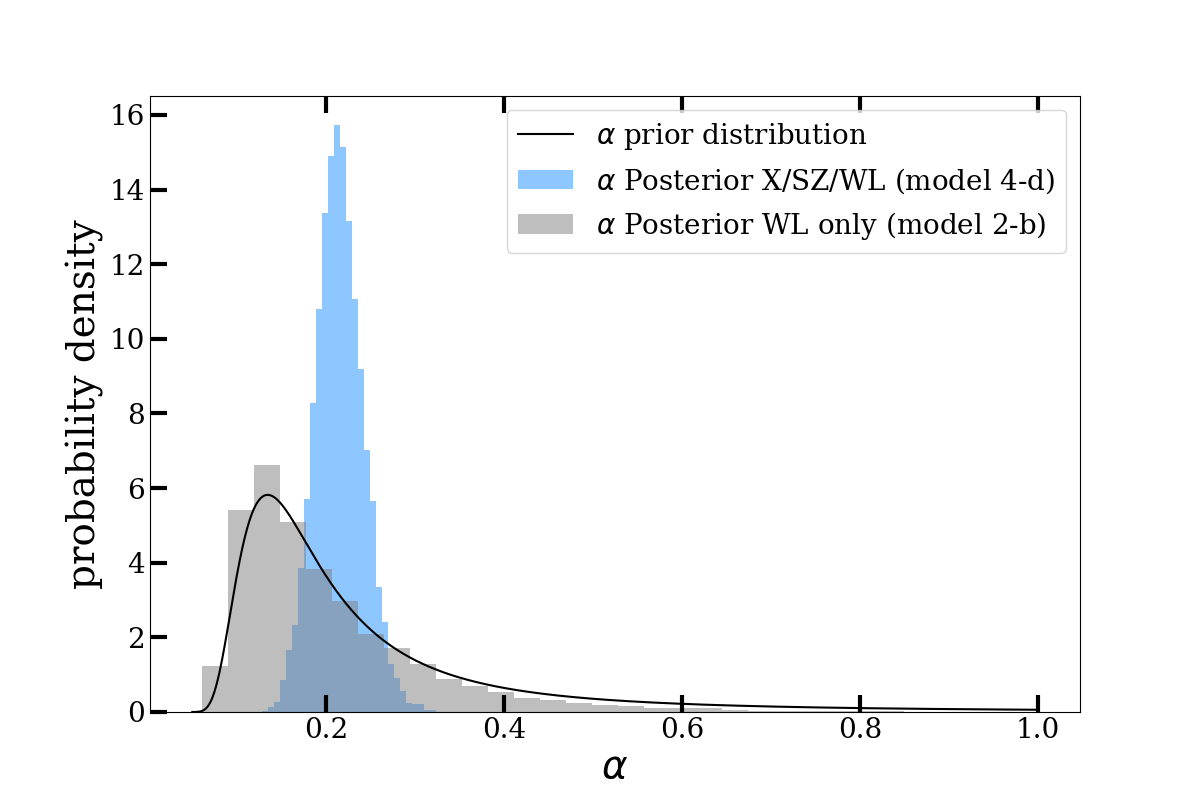}
    \caption{Einasto shape parameter $\alpha$ prior distribution over-plotted with its posterior distributions in two cases: weak lensing only analysis (model 2-b) and full joint analysis (X-ray/SZ/Weak Lensing joint analysis with elongation and polytropic non-thermal pressure support, model 4-d).}
    \label{fig:alpha_prior_posterior}
\end{figure}

\subsection{Successes and limitations of the model}

The LOS elongation recovered thanks to our modeling shows very good agreement with \citet{chexmate_2023}, where a full triaxial modeling of the cluster was carried out. The authors showed that A1689 is highly elliptical, and almost perfectly aligned with the LOS. This configuration explains the lower concentrations and masses inferred by the gas analysis of A1689 with respect to the lensing analysis, which shows a high lensing power and a large Einstein radius. Indeed, previous strong lensing analyses \citep{Broadhurst_2005} measured $\theta_{E} \sim 50^{\prime\prime}$ for the Einstein radius, which we can use as an independent test for our reconstruction. The expected Einstein radius can be predicted by our model by computing the radius $\theta_E$ within which the mean enclosed surface mass density equals the critical surface mass density. Assuming a mean source redshift $z_s=2$, from our WL-only analysis (model 2-a) we estimate $\theta_{E} = 60^{+4}_{-4}$ arcseconds, whereas from the full model we infer $\theta_{E} = 47^{+3}_{-6}$ arcseconds (models 4-b and 4-c). Thus, the Einstein radius predicted by our model is in excellent agreement with the measured Einstein radius. Since it accounts only for LOS elongation, our model does not probe a full triaxial geometry. The projection of the cluster in the plane of the sky is assumed to be circular or circularized through other means (e.g. azimuthal median). Thanks to the almost perfect alignment between the major axis and the LOS, A1689 is well suited for this analysis, which leads to an excellent agreement between the results obtained with this technique and the results of a full triaxial analysis \citet{Morandi_2011,Sereno_2012,Umetsu_2015,chexmate_2023}. Our results show that the main results of a full triaxial analysis can be mostly replicated with the addition of a single elongation parameter, which makes the model more stable numerically. On the other hand, our implementation of the underlying mass model is fully numerical, such that our framework is well suited for the study of extensions to the NFW mass model such as the Einasto model (see Sect. \ref{sec:einasto}). 

As shown in Sect. \ref{sec:XWL} and \ref{sec:full} and Fig. \ref{fig:pntVSelong_poly:xszwl}, the addition of SZ data allows our framework to break the degeneracy between non-thermal pressure and LOS elongation.
To further test this, we used the posterior predictive distributions of models 4-b and 4-c (our full multi-probe and elongated modeling approaches, implementing the polytropic and Angelinelli parameterizations of non-thermal pressure, respectively) to generate synthetic realizations of observables. Specifically, we drew 4$\times$2 sets of profiles, each consisting of $S_X$, $T_X$, $y$, and $g_+$.
We then fit these synthetic profiles using two different setups: (i) an X-ray/Weak Lensing (X/WL) analysis, and (ii) a full X-ray/SZ/Weak Lensing (X/SZ/WL) analysis, where the latter incorporates SZ constraints. This allowed us to isolate the role of SZ data in breaking the degeneracy between elongation and non-thermal pressure.
Fig. \ref{fig:ppd} presents a comparison of the constraints on $\alpha_{NT}(\frac{r_{200}}{2})$ and $e$ obtained from both analyses. The recovered values from X/WL and X/SZ/WL are compared against the "true" posterior distributions of $\alpha_{NT}(\frac{r_{200}}{2})$ and $e$ inferred from models 4-b and 4-c. 
In the absence of SZ data, the recovered elongation is biased low due to the prior. Adding SZ data, however, enables a successful recovery of A1689's median posterior distributions for both the elongation parameter $e$ and non-thermal pressure $\alpha_{NT}(\frac{r_{200}}{2})$, reducing their errors by factors of 5.5 and 1.4, respectively.

Therefore, we rely on the availability of high-quality observations in X-rays, SZ and weak lensing. To the present day, the available multi-wavelength data on A1689 are unrivalled, which makes this system the ideal candidate to validate our framework. However, upcoming large SZ (ACT, SPT) and weak lensing surveys (e.g. Euclid, Rubin) will bring datasets of similar quality for an increasing number of systems, which will allow us to study the distribution of three-dimensional cluster shapes and non-thermal pressure in sizable cluster samples.

As demonstrated by comparing our joint X-ray/SZ and X-ray/SZ/WL analysis, the main source of constraint on the LOS elongation parameter $e$ in our model comes from the ratio between the SZ and X-ray pressure profiles \citep{De_Filippis_2005, Sereno_2006, Bourdin_2017, Kozmanyan_2019}. This $\frac{P_{SZ}}{P_X}$ ratio also depends on the Hubble constant, and scaling with $h^{0.5}$\citep{Kozmanyan_2019, Ettori_2020}. Considering $h$ values ranging from 0.68 to 0.73, this results in an additional 3.6$\%$ uncertainty on the elongation parameter $e$.

\citet{Lau_2011} demonstrated that the gas distribution traces the shape of the gravitational potential, which usually differs from that of the DM distribution. As a result, the ellipticity estimated by our model is that of the gravitational potential rather than that of the mass distribution. In a triaxial geometry, the gravitational potential is usually more spherical than the underlying mass distribution, as it results from the point-to-point potential averaged over the entire system \citep{Sereno_2007,Kawahara_2010}. As a result, our assumption that the gas and DM are co-aligned may be overly simplistic. We showed in our full X-ray/SZ/WL multi-probe analysis that adding the weak lensing data does not impact the recovered LOS elongation, but instead constrains the non-thermal pressure. In case the ellipticity of the DM is larger than that of the potential (and, hence, the gas), the shear profile should be more affected by LOS elongation than the corresponding gas observables (Eq. \ref{eq:elong_vs_spherical}), such that the total mass and, in turn, the non-thermal pressure should be slightly lower than reported in this paper. Future versions of our pipeline will include a self-consistent treatment of LOS elongation on the gas and the DM.

\section{Conclusion}

In this paper, we introduced a new framework to reconstruct the shape of galaxy cluster mass profiles and non-thermal pressure support from the combination of X-ray, SZ, and WL data. We implemented a weak lensing module to the already existing code \begin{tt}hydromass\end{tt} \citep{Eckert_2022a}, thus adding the possibility to fit tangential shear profiles jointly with X-ray (temperature and surface brightness profiles) and SZ (Compton $y$ profile) data. To account for possible differences between the three observables, we introduced a LOS elongation parameter $e$ and non-thermal pressure support using two parameterizations (polytropic and Angelinelli). 

Applying our multi-probe framework to Abell 1689, we successfully fitted a density profile (NFW and Einasto) and were able to reproduce the observed profiles of X-ray surface brightness and temperature, Compton-$y$ parameter, and mean tangential shear within a common framework. We showed that A1689 is significantly elongated along the LOS, with the fitted elongation $e$ beind significantly different from unity with an uncertainty of only a few percent ($e$ = 1.30$\pm$0.03 from the full multi-probe analysis, accounting for non-thermal pressure; models 4-b and 4-c). When fitting only X-ray and WL data, the contribution of non-thermal pressure and LOS elongation is highly degenerate. The degeneracy can be broken by including the SZ data (see Fig. \ref{fig:pntVSelong_poly:xszwl}). Using the full dataset, our framework is able to recover the non-thermal pressure support, which constitutes the main source of hydrostatic bias when comparing X-ray galaxy cluster masses to weak lensing masses. The non-thermal pressure profiles obtained from the two parameterizations introduced in this work (polytropic and Angelinelli, see Sect. \ref{sec:model_NTpressure}) are consistent with one another, and yield a substantial non-thermal pressure support of $\sim40\%$ at $r_{500}$.


Our forward modeling approach of all the available observables (X-ray surface brightness and temperatures, Compton $y$, and mean tangential shear profiles) is fully numerical. As a result, our framework is able to perform a similar reconstruction for any density profile of choice and move beyond the assumption of an NFW density profile. In the NFW case, we showed that the very high concentration inferred from WL data only under the assumption of spherical symmetry ($c_{200}\sim15$, see Sect. \ref{sec:wlonly}) comes down to the more reasonable value of $\sim7$ when including the gas observables into the model (see Table \ref{table:combined}). Fitting the data with the Einasto model instead of NFW, we showed that both models provide a good fit to the A1689 dataset. Our multi-probe framework is able to constrain the Einasto shape parameter $\alpha=0.21\pm0.03$, whereas in the WL-only analysis the parameter does not pull away from the prior. 

Application of our framework to a larger sample of galaxy clusters such as CHEX-MATE \citep{chexmate_2021}, for which we find that 18 clusters (including A1689) are covered by ACT DR6 and part of the AMALGAM2 sample, will allow us at the same time to determine the shape of DM mass profiles, constrain the three-dimensional shape of galaxy clusters, and estimate the impact of non-thermal pressure support across the population.


\begin{acknowledgements}
Based on observations obtained with XMM-Newton, an ESA science mission with instruments and contributions directly funded by ESA Member States and NASA. DE acknowledges support from the Swiss National Science Foundation (SNSF) under grant agreement 200021\_212576. SE, MR, FG, IB, VG, HB, FDL, and PM acknowledge the financial contribution from the contracts Prin-MUR 2022, supported by Next Generation EU (n.20227RNLY3 {\it The concordance cosmological model: stress-tests with galaxy clusters}), ASI-INAF Athena 2019-27-HH.0, ``Attivit\`a di Studio per la comunit\`a scientifica di Astrofisica delle Alte Energie e Fisica Astroparticellare'' (Accordo Attuativo ASI-INAF n. 2017-14-H.0). EP acknowledges the support of the French Agence Nationale de la Recherche (ANR), under grant ANR-22-CE31-0010 (project BATMAN). RC acknowledges financial support from the INAF grant 2023 “Testing the origin of giant radio halos with joint LOFAR” (1.05.23.05.11). BJM acknowledges support from STFC grant ST/V000454/1. MG acknowledges support from the ERC Consolidator Grant \textit{BlackHoleWeather} (101086804). GWP acknowledges long-term support from CNES, the French space agency. This research was supported by the International Space Science Institute (ISSI) in Bern, through ISSI International Team project \#565 ({\it Multi-Wavelength Studies of the Culmination of Structure Formation in the Universe}).
\end{acknowledgements}

\bibliographystyle{aa} 
\bibliography{biblio} 

\appendix

\section{Comparison to triaxial modeling of \cite{chexmate_2023}}
\label{sec:kim2024}

In \cite{chexmate_2023}, a 3D triaxial modeling of A1689 was fitted to 2D maps of the same data set of gas observables (X-ray and SZ), using parameterized electronic density and gas pressure profiles and without hydrostatic assumption. The authors used a similar data set and also inferred the line of sight elongation through comparison of X-ray and SZ data, they offer a very comparable output. 
The recovered elongation parameter $e_{||}$ of their modeling is defined as the ratio of the size of the ellipsoid along the LOS ($l_{los}$) to the major axis of
the projected ellipse in the plane of the sky ($l_{p}$):

\begin{equation}
    e_{||} = \frac{l_{los}}{l_{p}}
\end{equation}

To enable a direct comparison between the line-of-sight elongation parameter obtained from our circular symmetry model and the triaxial model used by \citet{chexmate_2023}, we convert their fitted triaxial parameters (the minor-to-major and intermediate-to-major axial ratios denoted $q1$ and $q2$, inclination angle $\theta$, and Euler angles $\phi$ and $\psi$) into a comparable elongation measure $e$. Here, $e$ is defined as the ratio of the line-of-sight length of the cluster $l_{los}$ to $l_{eff}$, the effective radius of a circle with an equivalent area to that of the projected ellipse in the plane of the sky:

\begin{eqnarray}
    l_{eff} = l_p \sqrt{q_p},\\
    e = \frac{l_{los}}{l_{eff}},
\end{eqnarray}

where $q_p$ is the minor-to-major axial ratio of the observed projected isophote. Its definition can be found in Eq. 3 of \cite{chexmate_2023}, along with an exhaustive description of all the geometrical parameters involved in its computation. 
This conversion allows for a consistent comparison between the elongation derived from the full triaxial modeling with the plane of the sky circular symmetry assumption we make in our work.
\cite{chexmate_2023} recovered a value $e_{||} = 1.24 \pm 0.03$ which we convert into $e = 1.37 \pm 0.03$.


\section{Additional diagnostic plots}

\begin{figure*}[htbp]
  \begin{adjustwidth}{}{}
    \centering
    \begin{subfigure}[b]{0.48\textwidth}
      \centering
      \includegraphics[width=\linewidth, trim={0cm 0cm 0cm 0cm},clip]{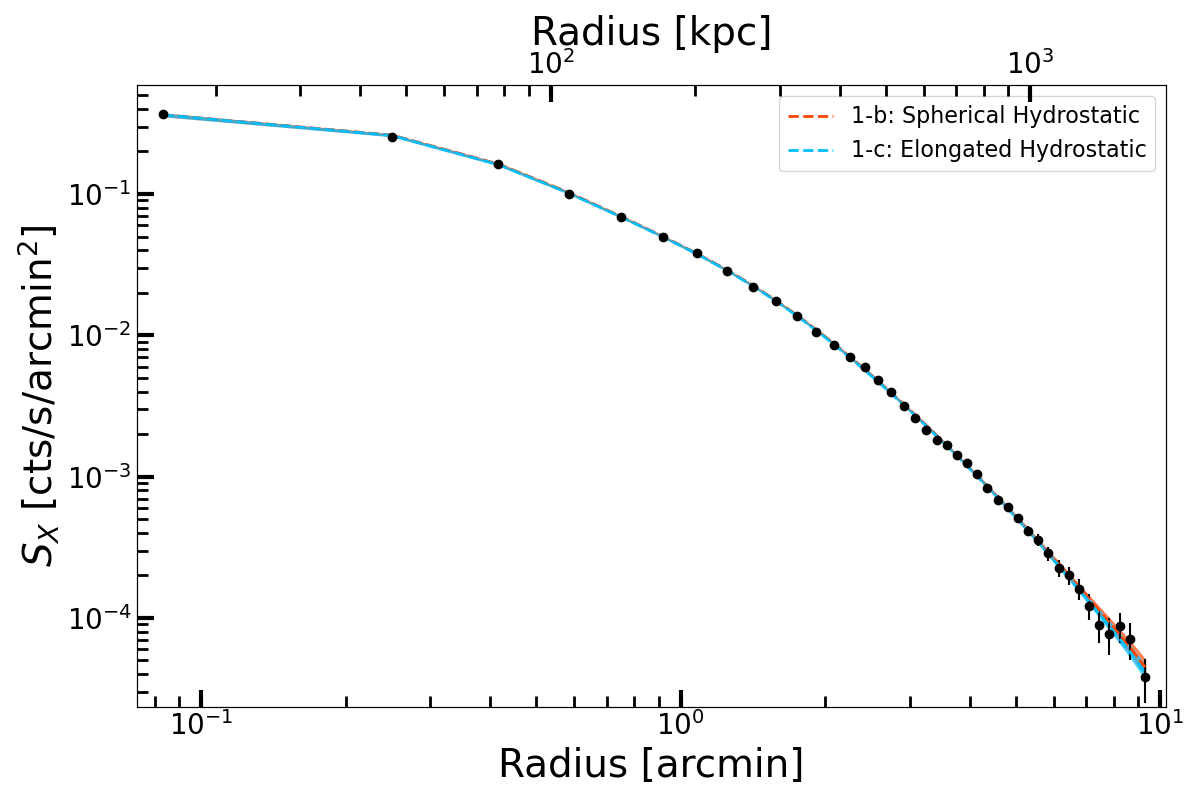}
      \caption{}
      \label{ap:xsz:sx}
    \end{subfigure}
    \hfill
    \begin{subfigure}[b]{0.48\textwidth}
      \centering
      \includegraphics[width=\linewidth, trim={0cm 0cm 0cm 0cm},clip]{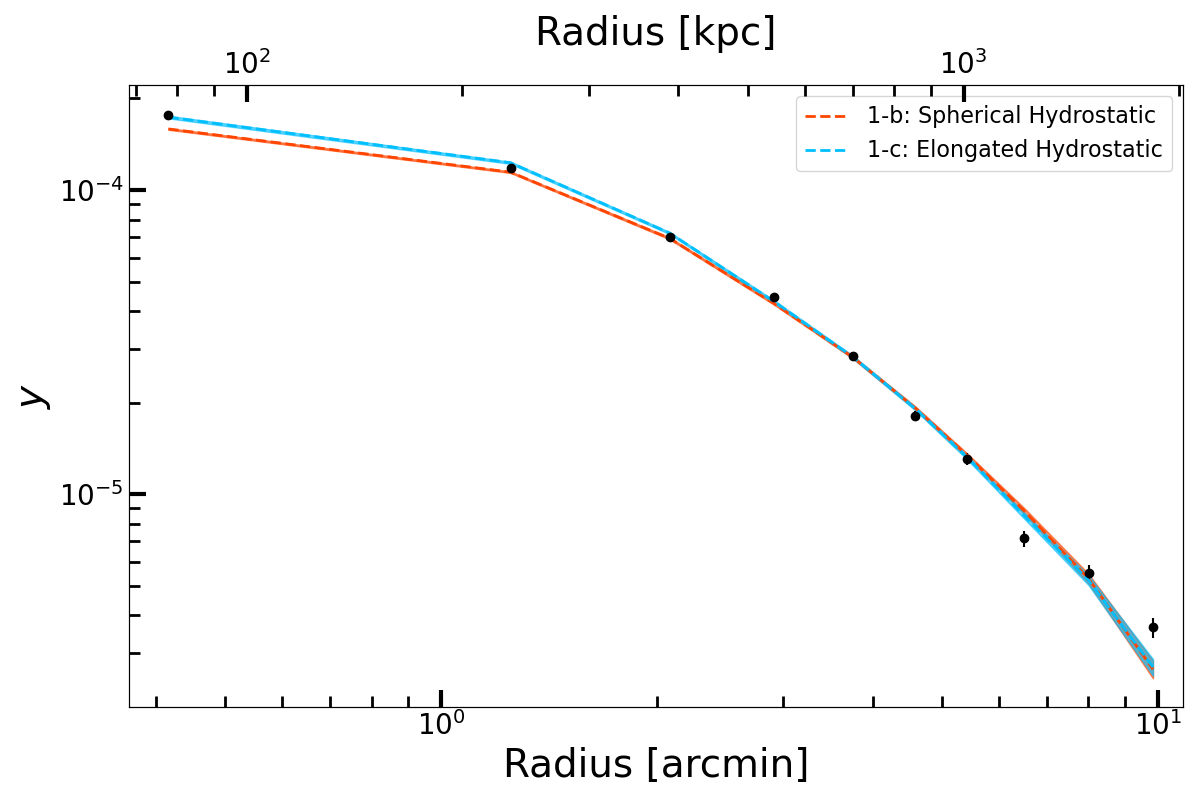}
      \caption{}
      \label{ap:xsz:y}
    \end{subfigure}
    \caption{A1689 X-ray/SZ, NFW joint analysis: fits to the X-ray brightness (left) and SZ $y$ parameter (right).}
    
    \label{ap:xsz}
  \end{adjustwidth}
\end{figure*}

\begin{figure*}[htbp]
  \begin{adjustwidth}{}{}
    \centering
    \begin{subfigure}[b]{0.45\textwidth}
      \centering
      \includegraphics[width=\linewidth, trim={0cm 0cm 0cm 0cm},clip]{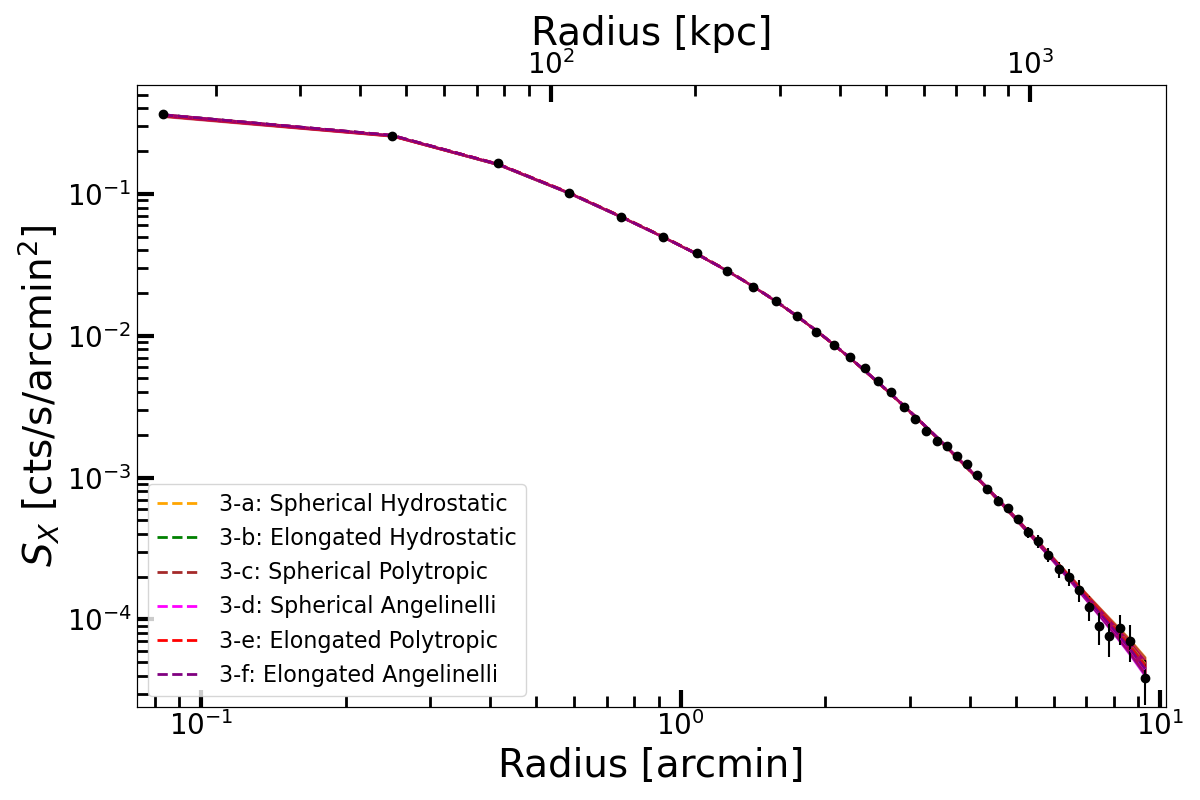}
      \caption{}
      \label{ap:xwl:sx}
    \end{subfigure}
    \hfill
    \begin{subfigure}[b]{0.45\textwidth}
      \centering
      \includegraphics[width=\linewidth, trim={0cm 0cm 0cm 0cm},clip]{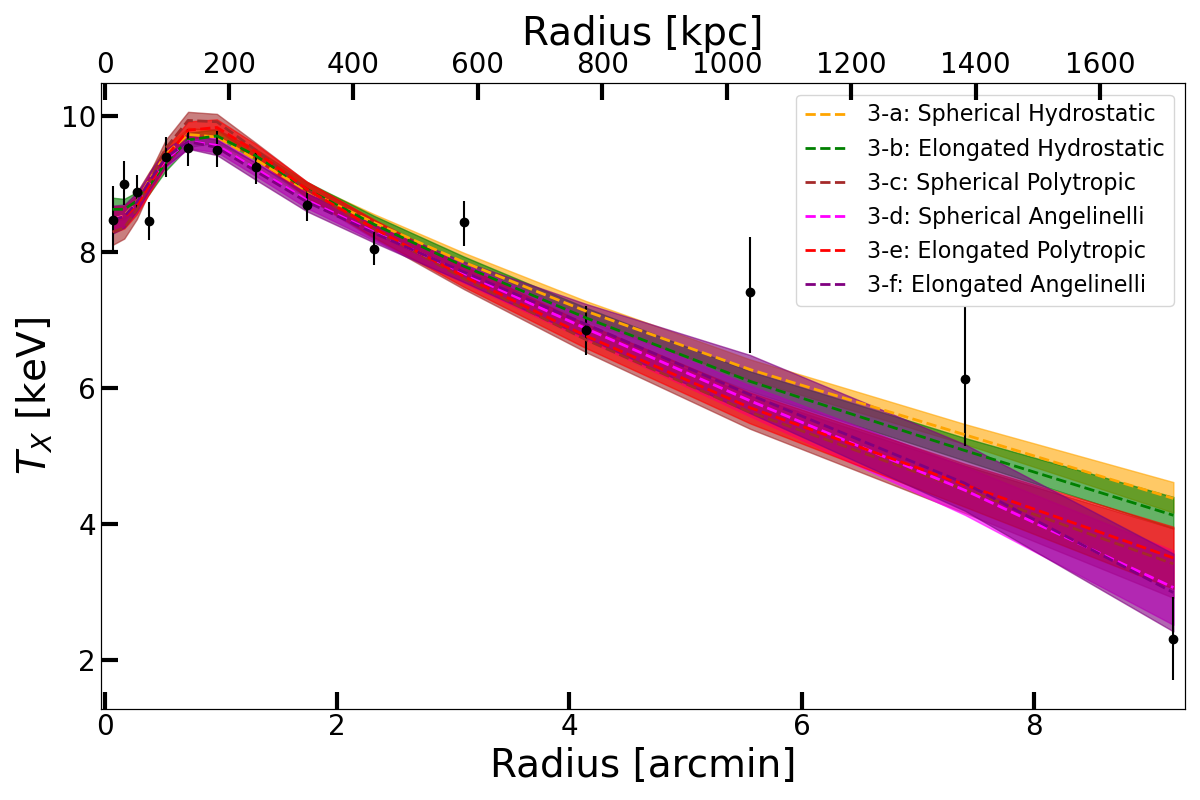}
      \caption{}
      \label{ap:xwl:tx}
    \end{subfigure}
    \caption{A1689 X-ray/WL, NFW joint analysis: fits to the X-ray observables.}
    
    \label{ap:xwl}
  \end{adjustwidth}
\end{figure*}

\begin{figure*}[htbp]
    \centering
    \begin{subfigure}[b]{0.48\textwidth} 
        \centering
        \includegraphics[width=\linewidth]{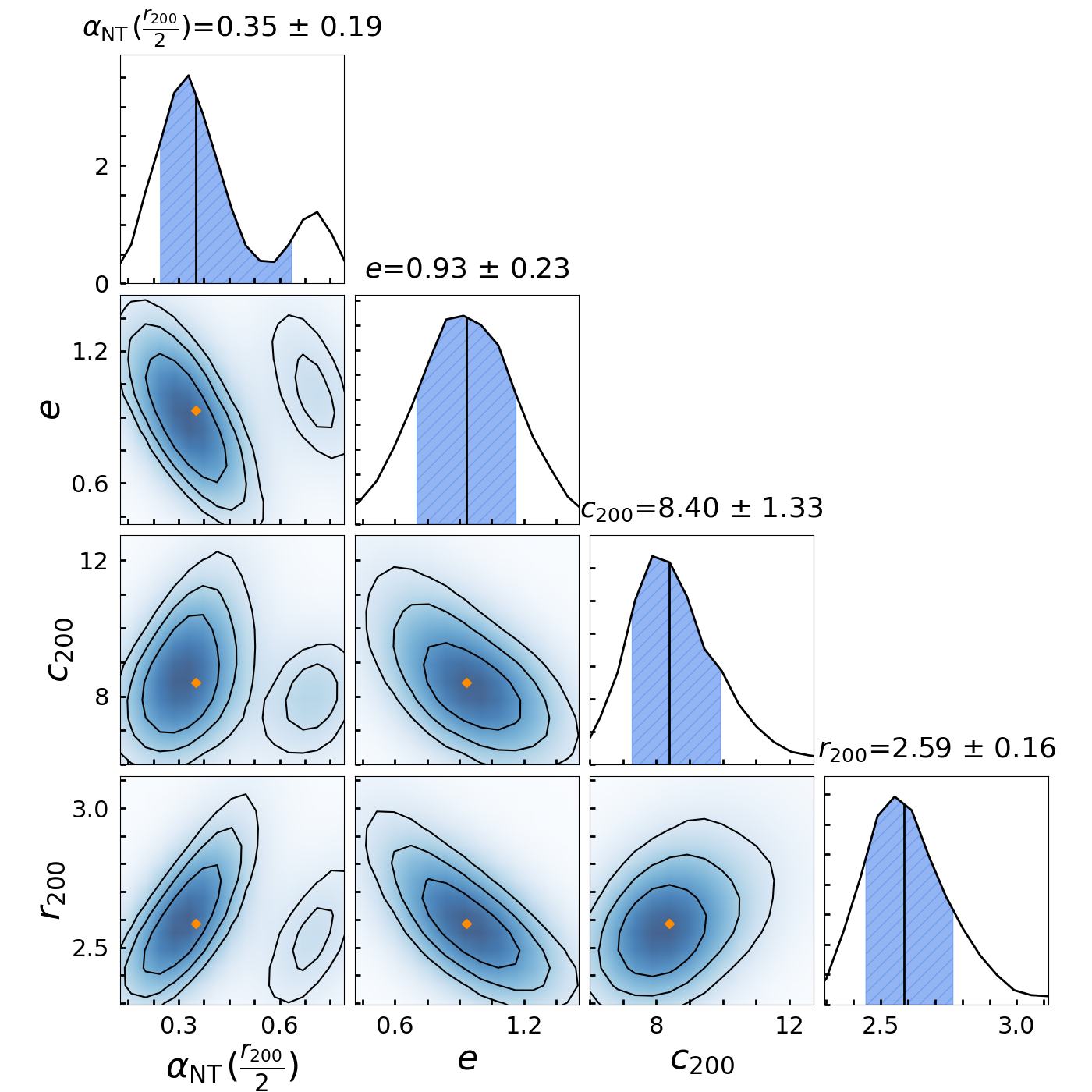}
        \caption{}
        \label{ap:pntVSelong_ang:xwl}
    \end{subfigure}
    \hfill
    \begin{subfigure}[b]{0.48\textwidth} 
        \centering
        \includegraphics[width=\linewidth]{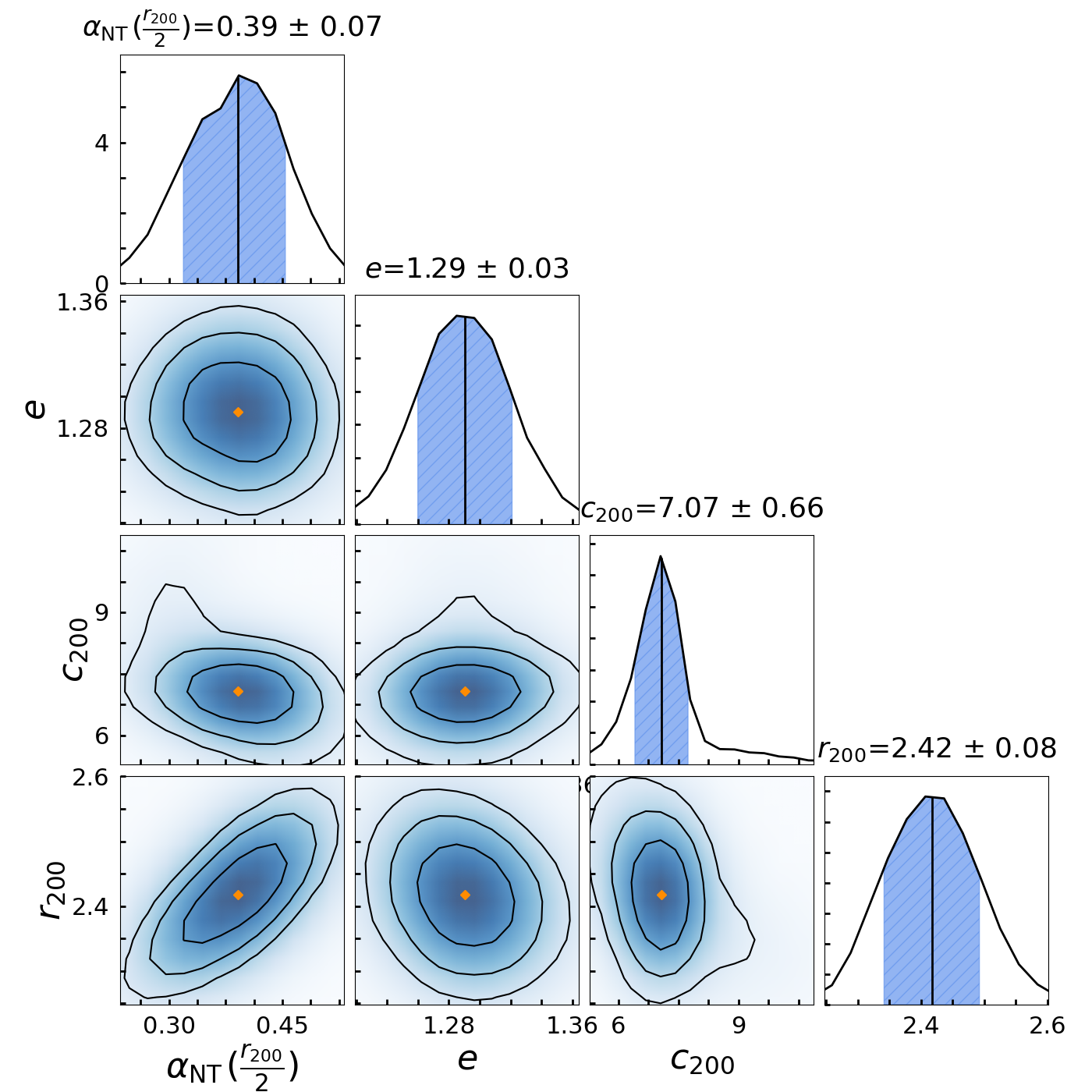}
        \caption{}
        \label{ap:pntVSelong_ang:xszwl}
    \end{subfigure}
    
    \caption{NFW joint analysis of A1689: corner plots featuring Angelinelli Non-thermal pressure to total pressure ratio $\alpha_{\operatorname{NT}}(0.5r_{200})$ , los elongation $e$, concentration $c_{200}$ and radius $r_{200}$ (in Mpc). \\
    a) X-ray/WL joint analysis,\\
    b) full X-ray/SZ/WL joint analysis.}
    \label{ap:pntVSelong_ang}
\end{figure*}


\begin{figure*}[htbp]
  \begin{adjustwidth}{}{}
    \centering
    \begin{subfigure}[b]{0.49\textwidth}
      \centering
      \includegraphics[width=\linewidth, trim={0cm 0cm 0cm 0cm},clip]{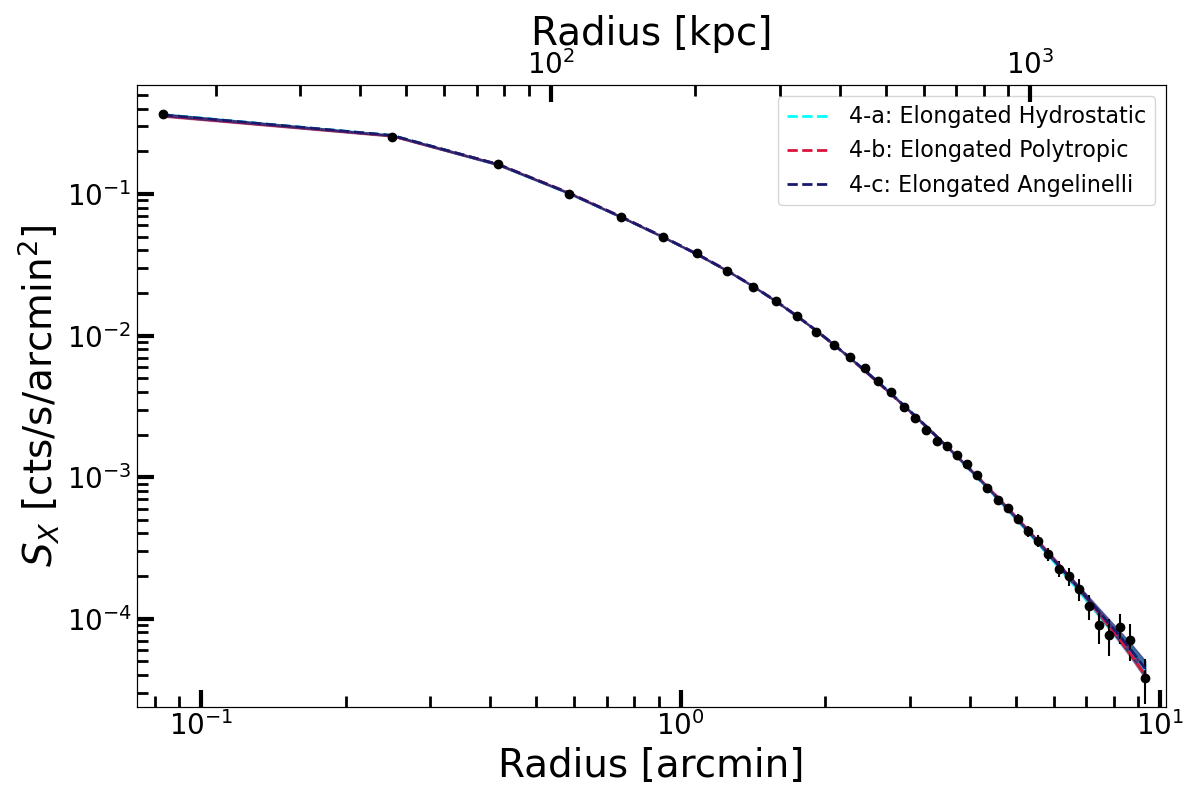}
      \caption{}
      \label{ap:xszwl:sx}
    \end{subfigure}
    \hfill
    \begin{subfigure}[b]{0.49\textwidth}
      \centering
      \includegraphics[width=\linewidth, trim={0cm 0cm 0cm 0cm},clip]{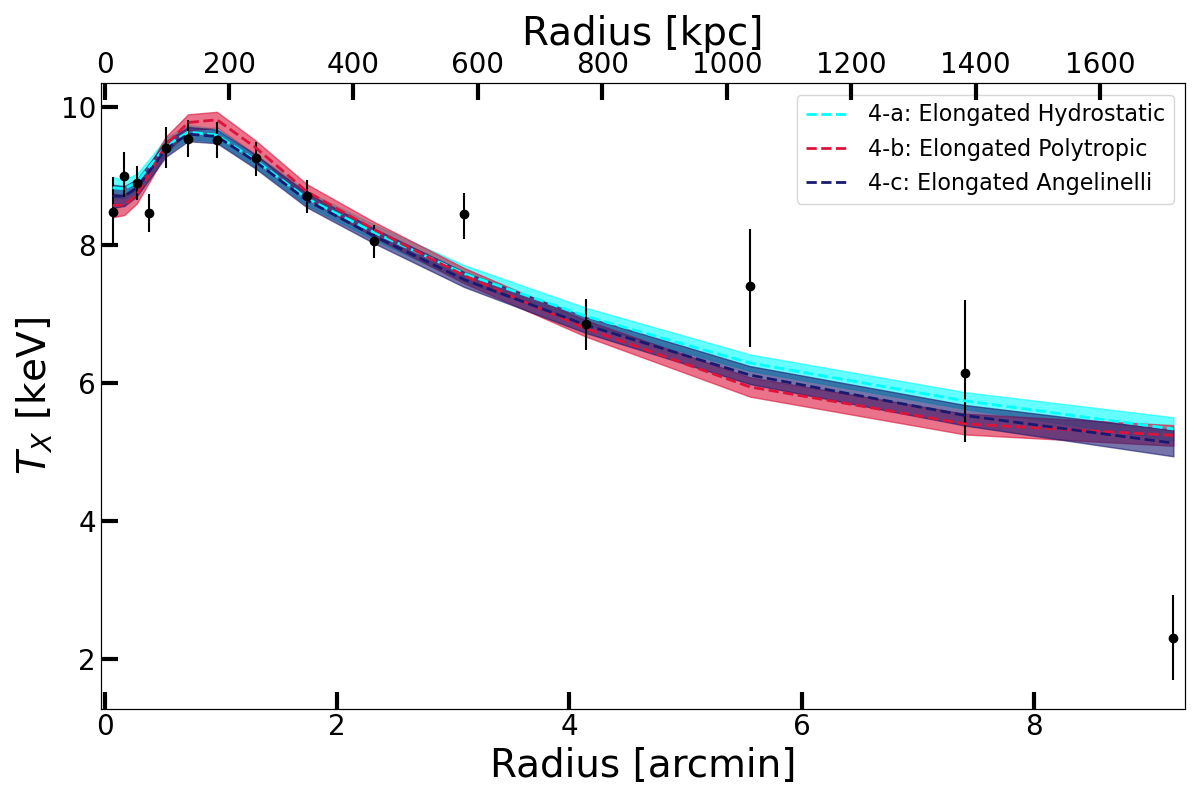}
      \caption{}
      \label{ap:xszwl:tx}
    \end{subfigure}
    \hfill
    \begin{subfigure}[b]{0.49\textwidth}
      \centering
      \includegraphics[width=\linewidth, trim={0cm 0cm 0cm 0cm},clip]{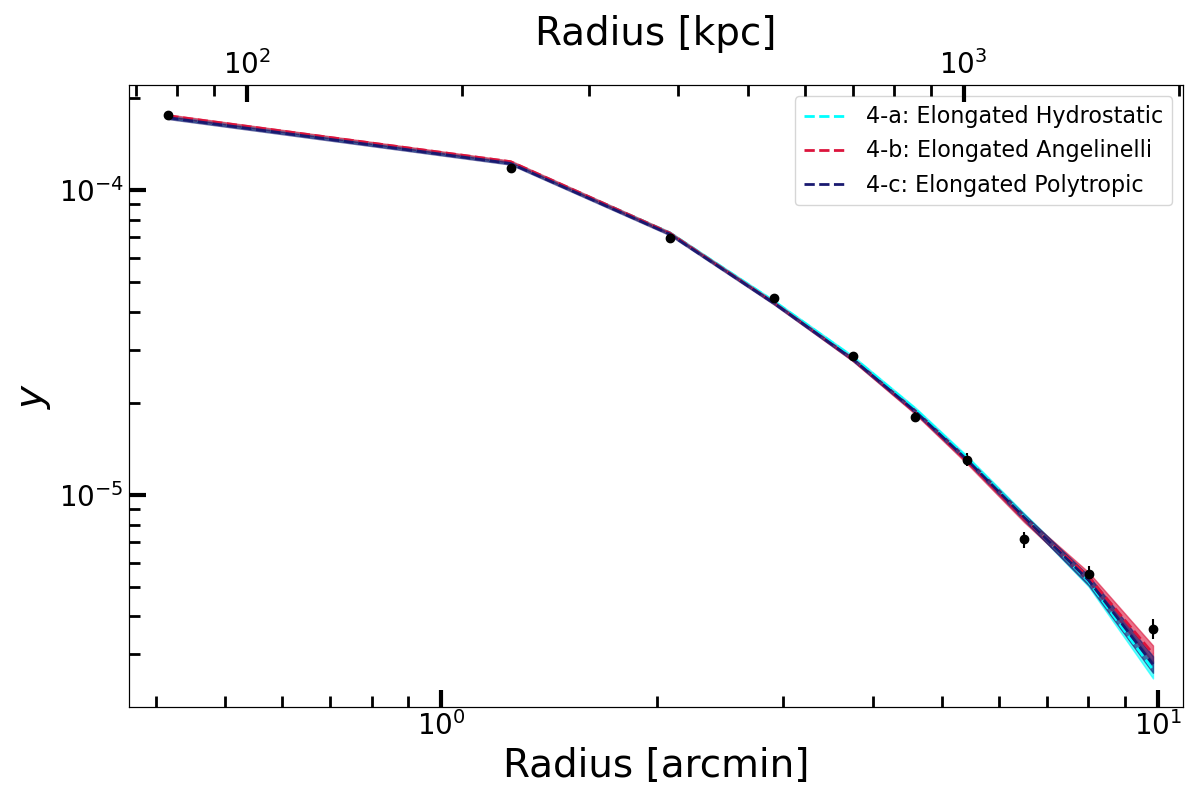}
      \caption{}
      \label{ap:xszwl:y}
    \end{subfigure}
    
    \caption{A1689 X-ray/SZ/WL, NFW joint analysis results: fits to the gas observables. }
    \label{ap:xszwl}
  \end{adjustwidth}
\end{figure*}


\begin{figure*}[htbp]
  \begin{adjustwidth}{}{}
    \centering
    \begin{subfigure}[b]{0.49\textwidth}
      \centering
      \includegraphics[width=\linewidth, trim={0cm 0cm 0cm 0cm},clip]{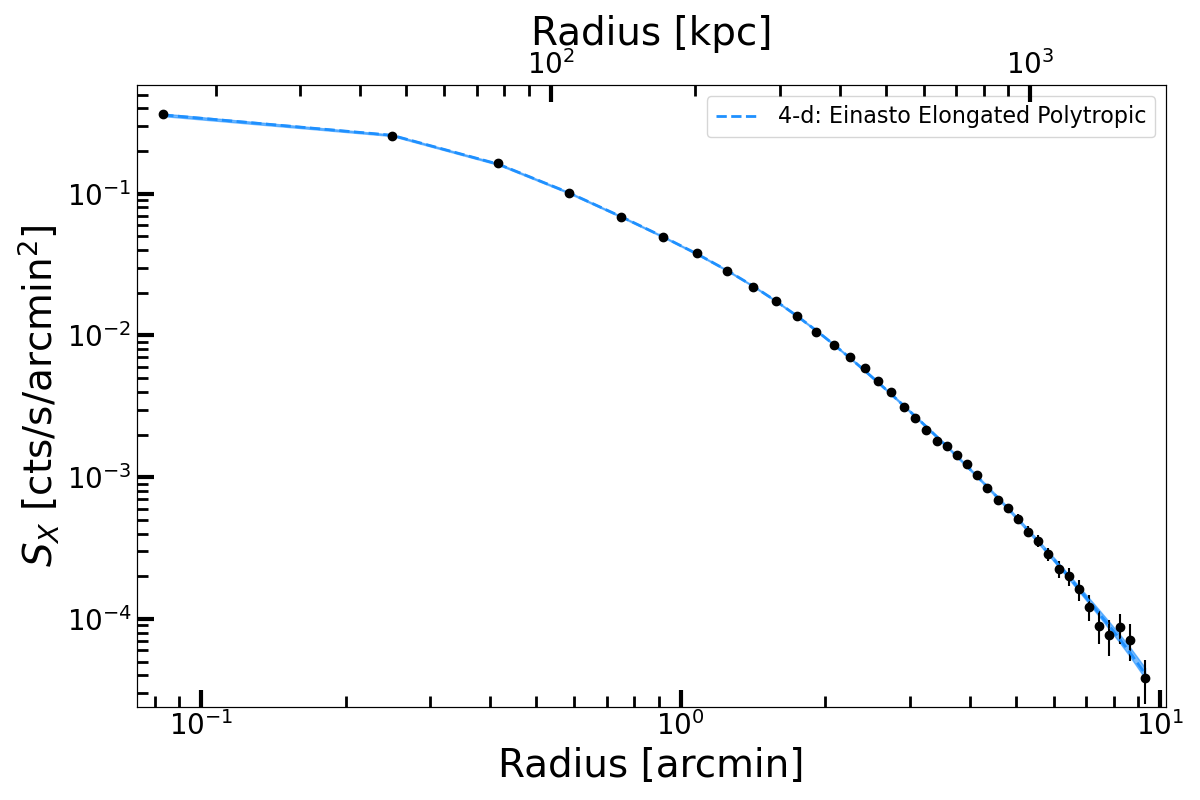}
      \caption{}
      \label{ap:einasto:sx}
    \end{subfigure}
    \hfill
    \begin{subfigure}[b]{0.49\textwidth}
      \centering
      \includegraphics[width=\linewidth, trim={0cm 0cm 0cm 0cm},clip]{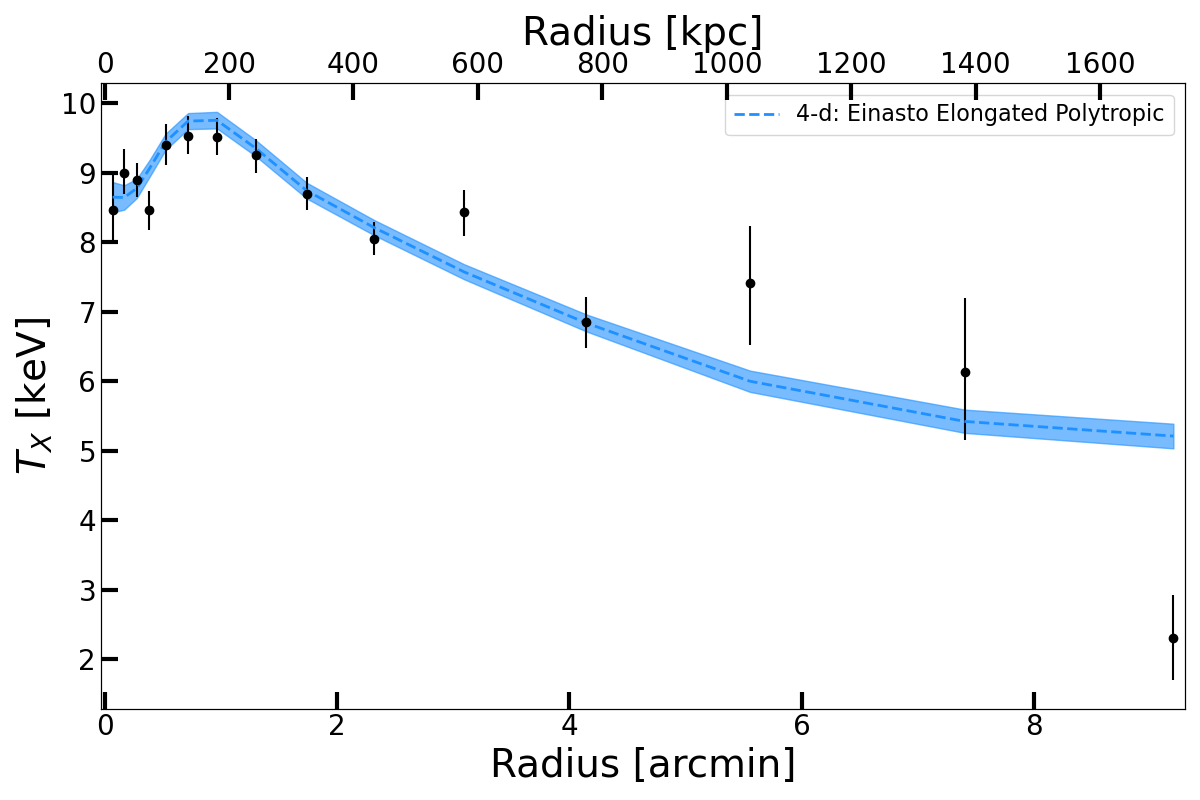}
      \caption{}
      \label{ap:einasto:tx}
    \end{subfigure}
    \hfill
    \begin{subfigure}[b]{0.49\textwidth}
      \centering
      \includegraphics[width=\linewidth, trim={0cm 0cm 0cm 0cm},clip]{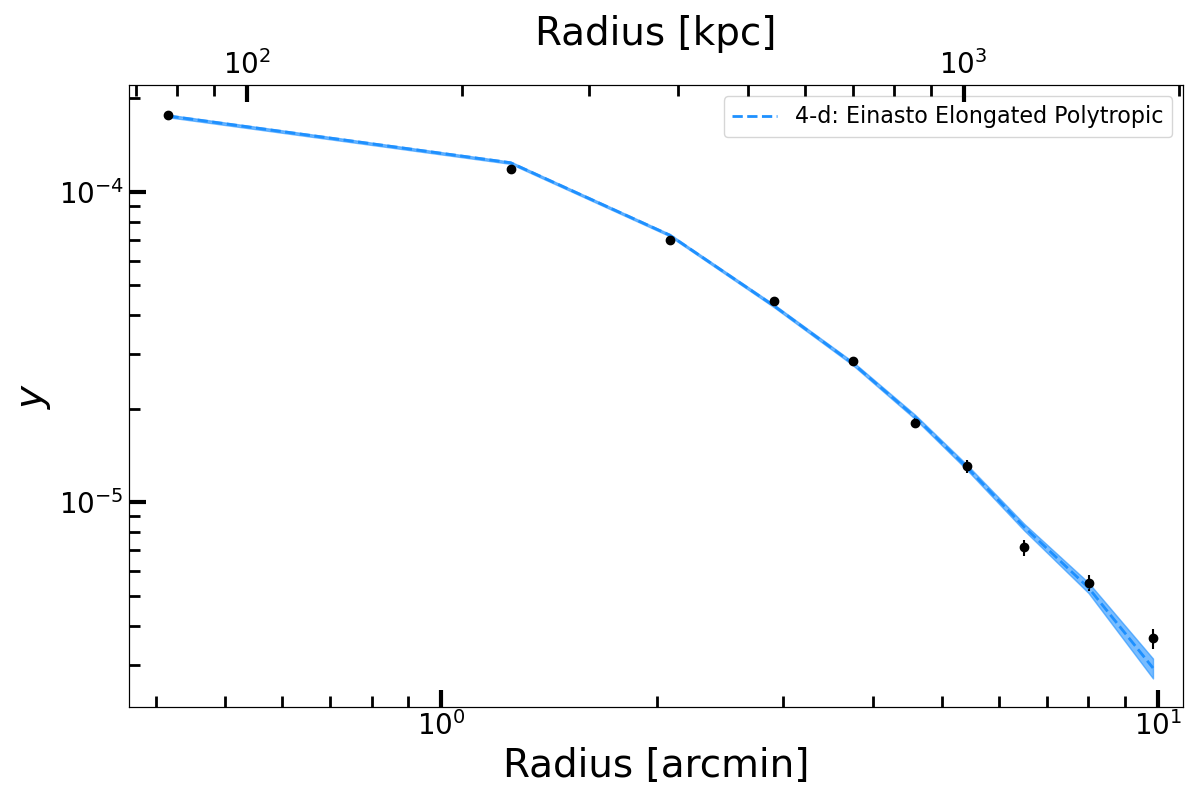}
      \caption{}
      \label{ap:einasto:y}
    \end{subfigure}
    \hfill
    \begin{subfigure}[b]{0.49\textwidth}
      \centering
      \includegraphics[width=\linewidth, trim={0cm 0cm 0cm 0cm},clip]{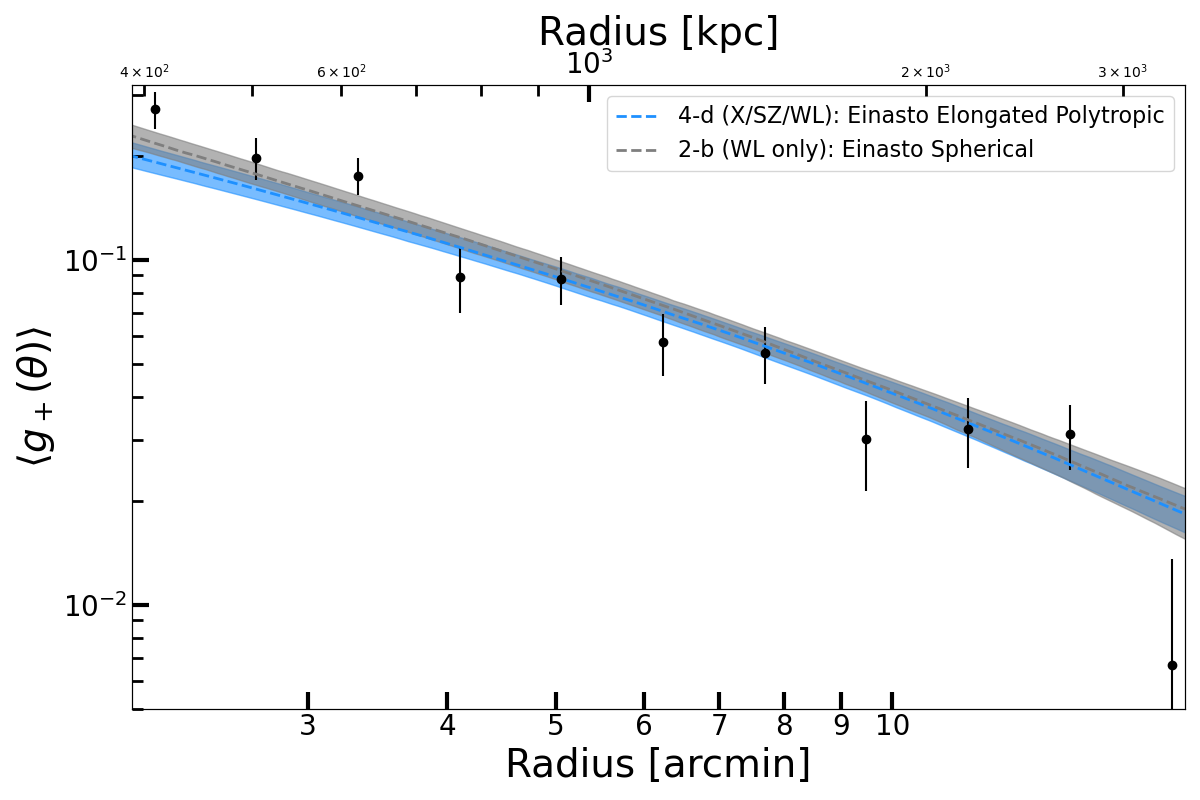}
      \caption{}
      \label{ap:einasto:gt}
    \end{subfigure}
    
    \caption{X-ray/SZ/WL joint analysis results: total A1689 data set fitted with the Einasto density profile in the elongated case with polytropic non-thermal pressure. The mean tangential shear profile also features the result for the Weak Lensing only Einasto analysis.}
    \label{ap:einasto}
  \end{adjustwidth}
\end{figure*}


\begin{figure*}[htbp]
    \centering
    \begin{subfigure}[b]{\linewidth}
      \centering
      \includegraphics[width=0.9\linewidth, trim={0cm 0cm 0cm 0cm},clip]{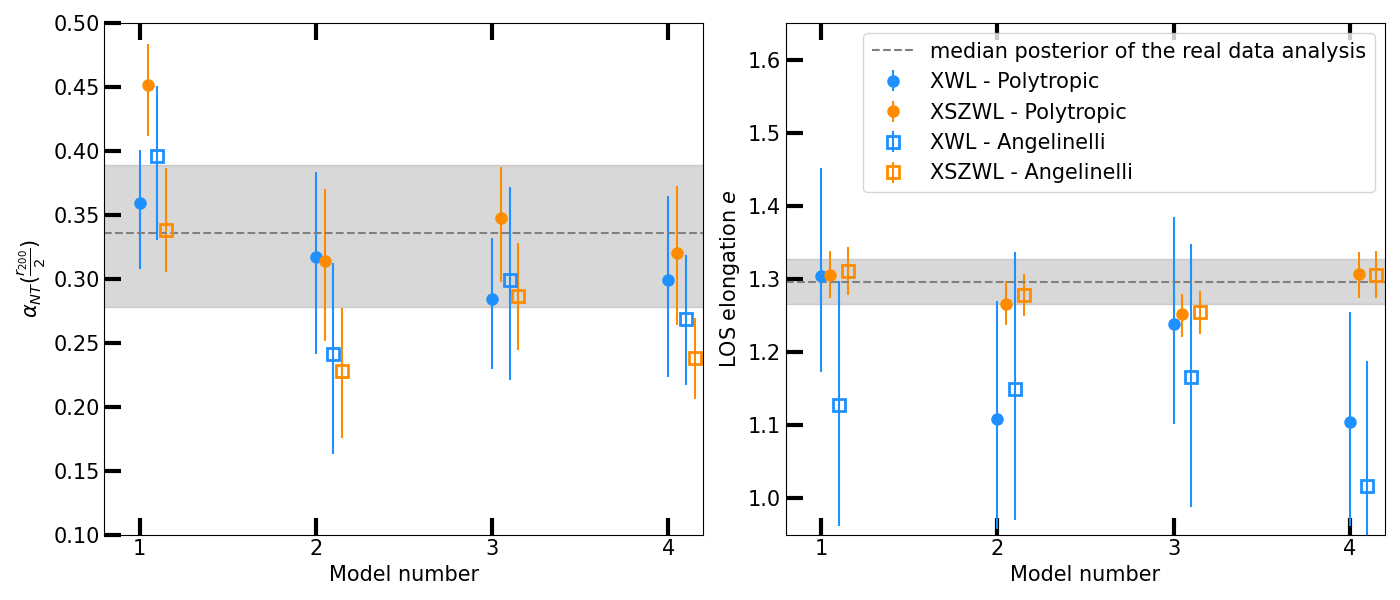}
      \caption{Analysis of mock profile drawn from model 4-b (polytropic) posterior predictive}
      \label{fig:ppd:ppd_poly}
    \end{subfigure}
    \hfill
    \begin{subfigure}[b]{\linewidth}
      \centering
      \includegraphics[width=0.9\linewidth, trim={0cm 0cm 0cm 0cm},clip]{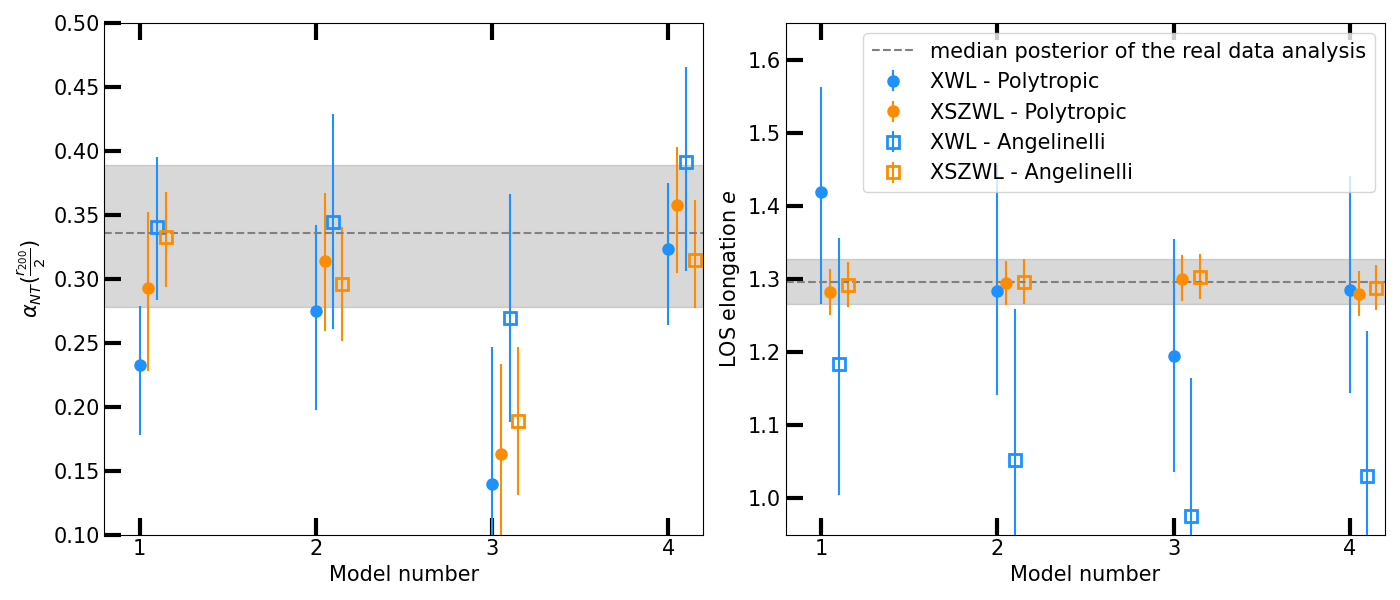}
      \caption{Analysis of mock profile drawn from model 4-c (Angelinelli) posterior predictive}
      \label{fig:ppd:ppd_ang}
    \end{subfigure}
    \caption{
    Comparison of the recovered values for the non-thermal pressure fraction, $\alpha_{NT}(\frac{r_{200}}{2})$, and elongation, $e$, obtained by fitting 4$\times$2 sets of profiles drawn from the posterior predictive distributions of our multi-probe analysis of A1689.
    The top panel presents results from profiles generated using model 4-b, which implements the polytropic parameterization. These profiles were fit using both the polytropic (circles) and Angelinelli (squares) parameterizations in two analysis setups: X-ray/Weak Lensing (blue) and X-ray/SZ/Weak Lensing (orange). The bottom panel shows the same procedure applied to profiles drawn from model 4-c, which follows the Angelinelli parameterization.
    The grey dashed lines and shaded regions indicate the median and 16th–84th percentile range of the posterior distributions for $\alpha_{NT}(\frac{r_{200}}{2})$ and $e$ in models 4-b (top) and 4-c (bottom). These serve as reference values, representing the expected constraints from the full multi-probe analysis, against which the recovered values from the fits can be compared.}
    
    \label{fig:ppd}
\end{figure*}

\end{document}